\documentclass[twocolumn]{article}

\usepackage{preprint}

\usepackage[utf8]{inputenc}     
\usepackage[T1]{fontenc}	    
\usepackage{microtype}	     	

\usepackage{abstract}           
\RequirePackage{algorithm}      
\RequirePackage[noend]{algpseudocode}
\usepackage{amsmath}
\usepackage{bm}                 
\usepackage{caption}            
\usepackage{mathtools}          
\usepackage{todonotes}

\makeatletter
\renewcommand{\ALG@beginalgorithmic}{\small}
\makeatother

\usepackage{siunitx}			
\DeclareSIUnit\angstrom{\text {\r{A}}}  
\DeclareSIUnit\cal{cal}
\DeclareSIUnit\kcal{\kilo\cal}

\usepackage{graphicx}
\graphicspath{ {./figures/} }

\usepackage[backend=biber,sorting=none,giveninits,style=numeric-comp,doi=false,isbn=false,url=false,date=year,eprint=false,maxnames=5]{biblatex}
\renewbibmacro{in:}{}
\AtEveryBibitem{\clearfield{pages}}
\newcommand{\abs}[1]{\left\lvert#1\right\rvert}
\newcommand{\mean}[1]{\langle#1\rangle}

\newcommand{\kT}{k_{\mathrm{B}}T}   
\newcommand{\Jm}[2][\map]{\abs{J_{#1}(#2)}} 
\newcommand{\logJm}[2][\map]{\log\Jm[#1]{#2}} 
\newcommand{\map}{\mathcal{M}}      
\newcommand{\iref}{1}               
\newcommand{\itarget}{2}            
\newcommand{\sref}{_{\iref}}      
\newcommand{\starget}{_{\itarget}}    
\newcommand{\sdiff}{_{\iref\itarget}}  
\newcommand{\x}{\bm{x}}             
\newcommand{\y}{\bm{y}}             
\newcommand{\w}[2][\map]{w[#1](#2)}

\title{Multimap targeted free energy estimation}

\date{}  

\usepackage{authblk}

\author[1,2\thanks{\tt{a.rizzi@fz-juelich.de}}]{Andrea Rizzi\orcid{0000-0001-7693-2013}}
\author[1,3\thanks{\tt{p.carloni@fz-juelich.de}}]{Paolo Carloni\orcid{0000-0002-9010-0149}}
\author[2\thanks{\tt{michele.parinello@iit.it}}]{Michele Parrinello\orcid{0000-0001-6550-3272}}
\affil[1]{Computational Biomedicine, Institute of Advanced Simulations IAS-5/Institute for Neuroscience and Medicine INM-9, Forschungszentrum J{\"u}lich GmbH, J{\"u}lich 52428, Germany}
\affil[2]{Atomistic Simulations, Italian Institute of Technology, Via Morego 30, Genova 16163, Italy}
\affil[3]{Department of Physics and Universit{\"a}tsklinikum, RWTH Aachen University, Aachen 52074, Germany}

\begin{document}

\twocolumn[
\maketitle
\begin{onecolabstract}
    We present a new method to compute free energies at a quantum mechanical (QM) level of theory from molecular simulations using cheap reference potential energy functions, such as force fields.
    To overcome the poor overlap between the reference and target distributions, we generalize targeted free energy perturbation (TFEP) to employ multiple configuration maps.
    While TFEP maps have been obtained before from an expensive training of a normalizing flow neural network (NN), our multimap estimator allows us to use the same set of QM calculations to both optimize the maps and estimate the free energy, thus removing almost completely the overhead due to training.
    A multimap extension of the multistate Bennett acceptance ratio estimator is also derived for cases where samples from two or more states are available.
    Furthermore, we propose a one-epoch learning policy that can be used to efficiently avoid overfitting when computing the loss function is expensive compared to generating data.
    Finally, we show how our multimap approach can be combined with enhanced sampling strategies to overcome the pervasive problem of poor convergence due to slow degrees of freedom.
    We test our method on the HiPen dataset of drug-like molecules and fragments, and we show that it can accelerate the calculation of the free energy difference of switching from a force field to a DFTB3 potential by about 3 orders of magnitude compared to standard FEP and by a factor of about 8 compared to previously published nonequilibrium calculations.
\end{onecolabstract}
\vspace{0.9cm}
]
\saythanks

\section{Introduction}

The accurate prediction of free energy differences is a long-sought goal in molecular simulations with many important applications to biology, pharmacology, and material science.
In drug discovery, in particular, free energy calculations of ligand binding affinities can accelerate the development of drug leads~\cite{schindler2020large,stampelou2022dual,cournia2017relative}.
These calculations are most often based on empirical interatomic potential energy functions~\cite{parks2020d3r,amezcua2021sampl7}, such as the CHARMM~\cite{mackerell1998all} or AMBER~\cite{ponder2003force} biomolecule force fields (FFs), owing to the good compromise offered by FFs between sampling efficiency and accuracy~\cite{parks2020d3r,amezcua2021sampl7}.

In recent years, there has been an increasing interest in predicting binding affinities using first-principles-based hybrid quantum mechanics/molecular mechanics (QM/MM) potentials~\cite{olsson2017comparison,hudson2018force,wang2019host,hudson2022obtaining,rufa2020towards}.
These might extend the domain of applicability of these methods to challenging systems such as transition metal-based enzymes~\cite{song2020thermodynamics,riccardi2018metal,li2017metal} and cover most of the drug-like chemical space with no or minimal parameterization.
However, these calculations have been so far mostly limited to small binding model systems~\cite{olsson2017comparison,hudson2018force,wang2019host,hudson2022obtaining} due to the prohibitive amount of sampling required to converge the free energy estimates.

The reference potential method~\cite{gao1992absolute,muller1995ab} (sometimes called indirect or bookend approach) provides a strategy to mitigate this problem.
In this scheme, the free energy difference is first predicted using a cheap reference potential (e.g., a force field) with standard free energy methods (e.g., alchemical methods~\cite{mey2020best}, umbrella sampling~\cite{torrie1977nonphysical,souaille2001extension}, metadynamics~\cite{laio2002escaping,barducci2008well,invernizzi2020rethinking}).
Then, a free energy correction at the QM(/MM) level of theory is calculated with free energy perturbation (FEP)~\cite{zwanzig1954high} techniques only for the states of interest.
This strategy avoids the need of sampling the entire molecular process with the expensive potential, and it is highly parallelizable, which is increasingly important given the dramatic advances in distributed computing~\cite{schneider2022exascale}.

However, perturbative techniques often fail for two main reasons.
First, in the presence of (usually unknown) slow degrees of freedom, it can be extremely challenging to sample thoroughly from the reference Boltzmann distribution and explore all the metastable states with a non-negligible population~\cite{ross2020enhancing,kaus2015deal,wang2012achieving,mobley2007confine} (e.g., ligand and side-chains torsional states).
This is crucial since, if the sampling is incorrect, no estimator will converge in reasonable time to the correct free energy.
In this case, it becomes necessary to adopt enhanced sampling techniques (for a recent review, see~\cite{henin2022enhanced}), which typically accelerate the dynamics by either adding a biasing potential or force along a collective variable (CV) (e.g., metadynamics~\cite{laio2002escaping,barducci2008well,invernizzi2020rethinking}, adaptive biasing force~\cite{darve2001calculating,henin2004overcoming}) or by modifying a parameter of the ensemble or even the Hamiltonian (e.g., Hamiltonian replica exchange~\cite{sugita2000multidimensional}, REST~\cite{liu2005replica,wang2011replica}).
Second, when the perturbation from the reference to the target potential is large, converging the free energy correction might be slow and even less efficient than the highly expensive direct sampling with QM.
Many solutions have been proposed to attenuate the problem~\cite{konig2014multiscale,dybeck2016comparison,giese2022multireference,shen2016multiscale,chehaibou2019computing,bucko2020ab,olsson2017comparison,steinmann2018relative,wang2019host,herzog2022assessing,hudson2015use,wang2018predicting,li2018accelerated,shen2018molecular,hudson2018force,giese2019development,pan2019accelerated}.
These methods either attempt to build a cheaper model of the target potential using machine learning~\cite{shen2016multiscale,chehaibou2019computing,bucko2020ab,herzog2022assessing} or (re)parameterize the reference potential to reduce its distance from the target~\cite{hudson2018force,giese2019development,pan2019accelerated}, possibly using multiple reference potentials~\cite{konig2014multiscale,dybeck2016comparison,giese2022multireference}.
However, even with these improvements, when reference and target are too different, it can still be impossible to converge free energy estimates without sampling from the target and/or intermediate potentials using equilibrium~\cite{olsson2017comparison,steinmann2018relative,wang2019host,herzog2022assessing} or nonequilibrium protocols~\cite{hudson2015use,wang2018predicting}.
This is usually the most expensive part of the calculation, although parallelizable methods such as nonequilibrium switching can significantly reduce the required wall-clock time.

Recently, we proposed to solve the problem using a configurational mapping approach that only requires sampling from the reference potential~\cite{rizzi2021targeted}.
Configurational mapping strategies date back to Voter~\cite{voter1985monte}, who noticed that when the two distributions differ mainly in the location of the energy minimum, the convergence can be considerably improved by first transforming the original samples through a displacement vector.
The targeted free energy perturbation (TFEP) method~\cite{jarzynski2002targeted} later generalized this idea to configurational maps rather than simple displacements.
Targeted extensions were later also proposed for the Bennett Acceptance Ratio (BAR)~\cite{bennett1976efficient,hahn2009using} and multistate BAR (MBAR)~\cite{shirts2008statistically,paliwal2013multistate} estimators, which employ samples from two or more states (rather than a single reference state).
While elegant, these approaches require the identification of complex analytical configurational maps, which can be challenging~\cite{severance1995generalized,ytreberg2005peptide,tan2010efficient,moustafa2015very,schultz2016reformulation,schieber2018using,schieber2019configurational}.
Recently, Wirnsberger \textit{et al.}~\cite{wirnsberger2020targeted} proposed learning such configurational maps by training normalizing flow neural networks (NN)~\cite{papamakarios2021normalizing,wirnsberger2022normalizing} in a framework called learned free energy perturbation (LFEP).

In our previous work~\cite{rizzi2021targeted}, we showed that LFEP significantly improves the convergence of the perturbative step in reference potential applications.
However, this improvement comes at the cost of training a NN, which can become the bottleneck of the calculation due to two fundamental sources of inefficiency.
First, evaluating the loss function is expensive since it requires computing single-point energies and forces of the mapped configurations with the expensive target potential.
These single-point calculations are used to optimize the NN and then discarded as they cannot be used to estimate the free energy difference.
Second, overfitting can introduce significant bias in the free energy estimate~\cite{rizzi2021targeted}.
The standard solution in machine learning is to monitor the validation loss on a separate set, but this is undesirable due to the computational cost of the loss function.
To avoid this, one can calculate the free energy on a set that is independent of the one used for training.
However, this means that the training and evaluation steps compete for data.
Moreover, there is no simple way of determining \textit{a priori} when to stop training the map, and a much too long (short) training could be uselessly expensive (insufficient).

In this work, we address these challenges.
In particular, 1) we introduce an extension of targeted estimators employing multiple mapping functions that allows removing almost entirely the training overhead.
2) We propose a simple one-epoch training policy that can serve as an efficient alternative to a validation set when evaluating the loss function is expensive compared to generating new data.
3) To solve the fundamental problem of sampling in the presence of slow degrees of freedom, we show that our multimap approach can be straightforwardly combined with enhanced sampling strategies to accelerate the reference simulation.
We validate our approach by testing it on the ``HiPen'' dataset~\cite{kearns2019good} of drug-like molecules and fragments, and we show that it  quickly recovers the converged free energy differences computed with more expensive nonequilibrium methods~\cite{scholler2022optimizing} between CGenFF force field~\cite{vanommeslaeghe2010charmm} and the DFTB3 potential~\cite{elstner1998self,gaus2013parametrization}.
Finally, we demonstrate that the combined use of multimap TFEP and on-the-fly probability enhanced sampling (OPES)~\cite{invernizzi2020rethinking} can obtain converged free energy differences also for molecules displaying slowly-interconverting torsional states.

\section{Theory}

We begin this section by stating the problem and recalling the formulation of TFEP~\cite{jarzynski2002targeted}.
Next, we generalize the targeted estimators to incorporate multiple maps, and we propose a practical method based on normalizing flow neural networks.
We show that the method removes the need for a separate, expensive training phase, thus simplifying the application of TFEP and reducing considerably the cost of computing the free energy difference.
Finally, we introduce a one-epoch learning policy as a means to avoid overfitting.

\subsection{Targeted free energy perturbation}

\begin{figure}[t]
    \includegraphics[width=0.45\textwidth]{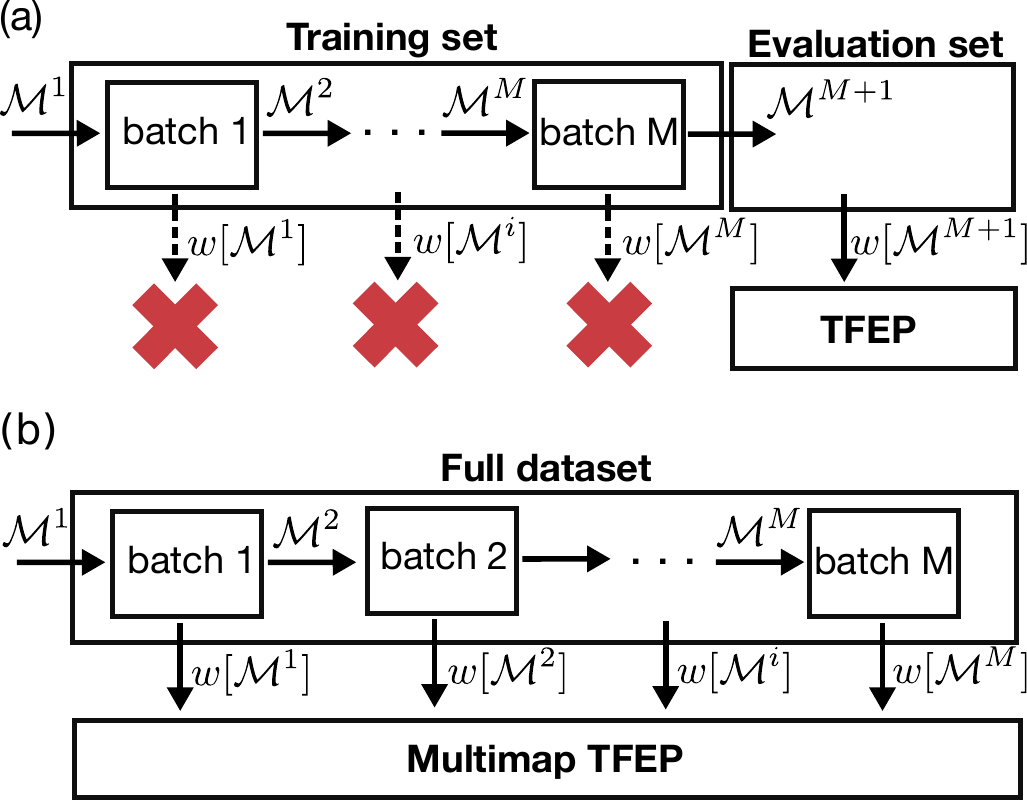}
    \centering
    \caption{\label{fig:mtfep}Schematics describing the protocols using (a) standard and (b) multimap TFEP.
    Multimap TFEP allows estimating the free energy using the work values generated during training, which are instead discarded in standard TFEP.}
\end{figure}

The main problem addressed in this article is the calculation of the free energy difference
\begin{equation}
	\Delta f\sdiff = f\starget - f\sref = - \log \frac{\int_{\Gamma\starget} e^{- u\starget(\y)} d\y}{\int_{\Gamma\sref} e^{- u\sref(\x)} d\x}
\label{eq:free-energy-diff}
\end{equation}
between a system described by a reference potential $u\sref(\x)$ and one described by a target potential $u\starget(\x)$.
In Eq.~\eqref{eq:free-energy-diff}, energies are defined in units of $\kT$, and $\Gamma_{\iref}$ and $\Gamma_{\itarget}$ are the domains of the reference and target Boltzmann distributions, respectively.

To derive the TFEP identity~\cite{jarzynski2002targeted}, we perform a change of variable $\y = \map(\x)$ in the numerator of Eq.~\eqref{eq:free-energy-diff}, where $\map : \Gamma\sref \to \Gamma\starget$ is invertible.
Then, by multiplying and dividing the integrand by $e^{u\sref(\x)}$ one obtains the result
\begin{equation}
    \Delta f\sdiff = - \log \mean{ e^{-\w{\x}} }\sref \; ,
\label{eq:tfep}
\end{equation}
where $\mean{}\sref$ represents the ensemble average over the reference distribution with potential $u\sref(\x)$, and we have defined the generalized work
\begin{equation}
\w{\x} = u\starget(\map(\x)) - \logJm{\x} - u\sref(\x) \; .
\label{eq:generalized-work}
\end{equation}
Eq.~\eqref{eq:tfep} gives us the means to compute $\Delta f\sdiff$ by running a simulation using only the reference potential.
The standard free energy perturbation (FEP) expression originally introduced by Zwanzig~\cite{zwanzig1954high} is recovered when $\map$ is set to the identity function.
The advantage of TFEP can be understood by realizing that the change of variable transforms the target potential into an effective potential of the form $u\starget'[\map](\x) = u\starget(\map(\x)) - \logJm{\x}$.
We can thus minimize the perturbation from $u\sref$ to $u\starget'[\map]$ by choosing a suitable $\map$, thus improving the convergence of the average calculated in Eq.~\eqref{eq:tfep}.

\subsection{Learned free energy perturbation}

To learn the configurational map by machine learning, we need a loss function to optimize~\cite{wirnsberger2020targeted,rizzi2021targeted}.
To this end, we define the Boltzmann distribution associated with the effective potential $p\starget'[\map](\x) = e^{f\starget - u\starget'[\map](\x)}$.
Our goal is to find $\map$ such as $p\starget'[\map](\x)$ is as close as possible to the reference Boltzmann distribution $p\sref(\x)$.
It is easy to show that, in the limiting case of $p\starget'[\map](\x) = p\sref(\x)$, the perturbation is effectively null, and the average in Eq.~\eqref{eq:tfep} converges with just one sample~\cite{jarzynski2002targeted}.
Thus, given $N$ samples from $p\sref(\x)$, we obtain the map by minimizing the negative log-likelihood of the data (up to an irrelevant constant $1/N$):
\begin{equation}
\begin{split}
    \mathcal{L} &= - \frac{1}{N} \log \prod_{i=1}^N p\starget'[\map](\x_i) \\
    &= \frac{1}{N} \sum_{i=1}^N u\starget(\map(\x_i)) - \logJm{\x_i} - f\starget \; .
\label{eq:tfep-loglikelihood}
\end{split}
\end{equation}
The term $f_2$ can be ignored during training since it is independent of $\map$.
In the limit of $N \to \infty$, Eq~\eqref{eq:tfep-loglikelihood} is equivalent to minimizing the KL divergence between $p\sref(\x)$ and $p\starget'(\x)$~\cite{rizzi2021targeted}.
Normalizing flow neural networks are ideally suited to this task as they are invertible by construction and enable the calculation of the Jacobian determinant in linear time~\cite{papamakarios2021normalizing}.

\subsection{Multimap free energy estimation}

\begin{algorithm}[!b]
\caption{Multimap TFEP with one-epoch learning}\label{alg:mtfep}
\begin{algorithmic}
\Require
    \State $N$ configurations $\bm{X} = [\x_1, \dots, \x_N]$ sampled from $p_1(\x)$
    \State batch size $L$  \Comment{Assume $N$ is divisible by $L$ for simplicity}
    \State
    \State $\map^1 \gets identity$  \Comment{First guess for the map}
    \State $M \gets N/L$  \Comment{Number of batches}
    \ForAll{$m \in \{1, \dots, M\}$}  \Comment{Loop over batches}
        \ForAll{$l \in \{1, \dots, L\}$}  \Comment{Loop over batch samples}
            \State $\x_{ml} \gets pick(\bm{X})$  \Comment{Sampling without replacement}
            \State $\y_{ml} \gets M^m(\x_{ml})$  \Comment{Map the configuration}
            \State $u_{ml} \gets u_2(\y_{ml})$  \Comment{Single-point QM calculation}
            \State $u'_{ml} \gets u_{ml} - \logJm[M^m]{\x_{ml}}$  \Comment{Effective potential}
            \State $w_{ml} \gets u'_{ml} - u\sref(\x_{ml})$  \Comment{Generalized work}
        \EndFor
        \State $\mathcal{L} \gets mean([u'_{m1}, \dots, u'_{mL}])$  \Comment{Loss function, Eq.~\eqref{eq:tfep-loglikelihood}}
        \State $\map^{m+1} \gets optimize(\mathcal{L}, \map^m)$  \Comment{Backpropagation}
    \EndFor
    \State $\Delta \hat{f}_{12} \gets - logmeanexp(-[w_{11}, \dots, w_{ML}])$  \Comment{Eq.~\eqref{eq:mtfep-estimator}}
\end{algorithmic}
\end{algorithm}

In learned FEP, all the single-point energy calculations performed during training are discarded after each NN optimization step because the loss in Eq.~\eqref{eq:tfep-loglikelihood} requires the potential of the mapped configuration, which is different in each training epoch.
As these calculations are typically very expensive in reference potential applications, this has a large negative impact on the overall efficiency of the method.
However, we note here that the problem would disappear if one could use the potentials generated during training to also compute the free energy estimate.
To this end, the main theoretical obstacle to overcome is that these data come from multiple distributions because the map changes at each optimization step.

Multimap free energy estimation solves this problem.
The critical observation is that Eq.~\eqref{eq:tfep} is valid for any invertible map $\map : \Gamma\sref \to \Gamma\starget$.
We can thus consider a collection of $M$ such maps $\{ \map^m \}_{m=1}^M$ and write
\begin{equation}
    \Delta f\sdiff = - \log \frac{1}{M} \sum_{m=1}^M \mean{ e^{-\w[\map^m]{\x}} }\sref \; .
\label{eq:mtfep}
\end{equation}

The collection of maps is completely arbitrary.
In this work, we generate a sequence of maps during batch training as shown in Fig.~\ref{fig:mtfep}.
We then estimate the free energy using
\begin{equation}
    \Delta \hat{f}\sdiff = - \log \frac{1}{ML} \sum_{m=1}^M \sum_{l=1}^L e^{-\w[\map^m]{\x_{ml}}} \; ,
\label{eq:mtfep-estimator}
\end{equation}
where $L$ is the batch size, $M$ is the total number of batches in the dataset, $\map^m$ is the map used to compute the loss in the $m$-th batch, and $\x_{ml}$ is the $l$-th sample within the $m$-th batch.

\begin{figure*}[!ht]
    \includegraphics[width=0.9\textwidth]{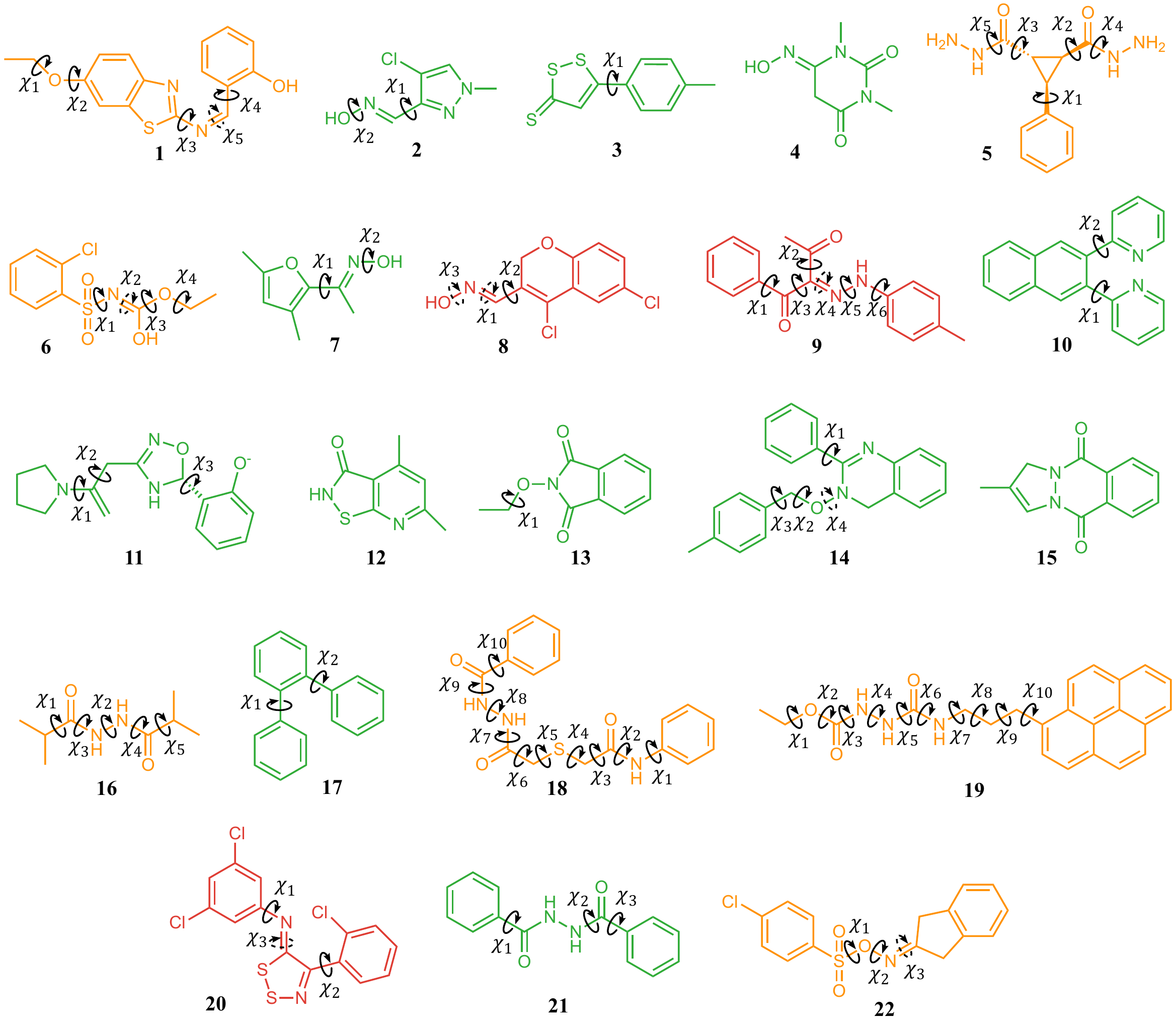}
    \centering
    \caption{\label{fig:hipen-dataset}The molecules in the HiPen dataset~\cite{kearns2019good} are classified into ``good" (green), ``bad" (orange), and ``ugly" (red).
    Solid arrows represent the dihedrals that were randomized to create the initial configurations for the replicate reference simulations.
    Dashed arrows indicate dihedrals that are discussed in the article but were not randomized.}
\end{figure*}

We stress that the summation in Eq.~\eqref{eq:mtfep-estimator} converges exactly to the free energy predicted by the target potential in the limit of an infinite number of independent samples for any collection of invertible maps generated during training.
To prove this, we cannot use the classical central limit theory because the random variables $e^{-\w[\map^m]{\x_{ml}}}$ entering the sum are not identically distributed due to the map changing during training.
Instead, we can invoke Kolmogorov's strong law~\cite[Chapter~X]{feller1968introduction}.
This states that (under suitable conditions described below) given a sequence of $N$ independent random variables $X_1, \dots, X_N$ with means $\mu_1, \dots, \mu_N$ and variances $\sigma_1, \dots, \sigma_N$, then
\begin{equation}
    \frac{1}{N} \left( \sum_{i=1}^N X_i - \sum_{i=1}^N \mu_i \right) \to 0 \; .
\label{eq:kolmogorov-strong-law}
\end{equation}
The convergence of Eq.~\eqref{eq:mtfep-estimator} follows by noticing that all elements of the summation have identical mean $\mu_i = e^{-\Delta f_{12}}$.
A sufficient condition for Kolmogorov's strong law to apply is that the variance of the summand in Eq.~\eqref{eq:mtfep-estimator} does not diverge  linearly or superlinearly as the training progresses, which is, in practice, easily satisfied unless the training of the NN becomes unstable (e.g., due to an excessive learning rate).
Note also that the result is valid even if the maps are correlated.
Standard and (single-map) TFEP are special cases of Eq.~\eqref{eq:mtfep-estimator} where the maps are constant.
It is straightforward to show that, similarly to standard FEP and TFEP~\cite{hahn2009using}, the estimate obtained in a finite time is positively biased, i.e., $\mean{ \Delta \hat{f}\sdiff } \ge \Delta f\sdiff$.

Similar considerations can be extended to the MBAR estimator~\cite{shirts2008statistically,paliwal2013multistate}, and its multimap generalization is derived in Appendix~\ref{sec:si:mtmbar}.

\begin{figure*}[!ht]
    \includegraphics[width=0.9\textwidth]{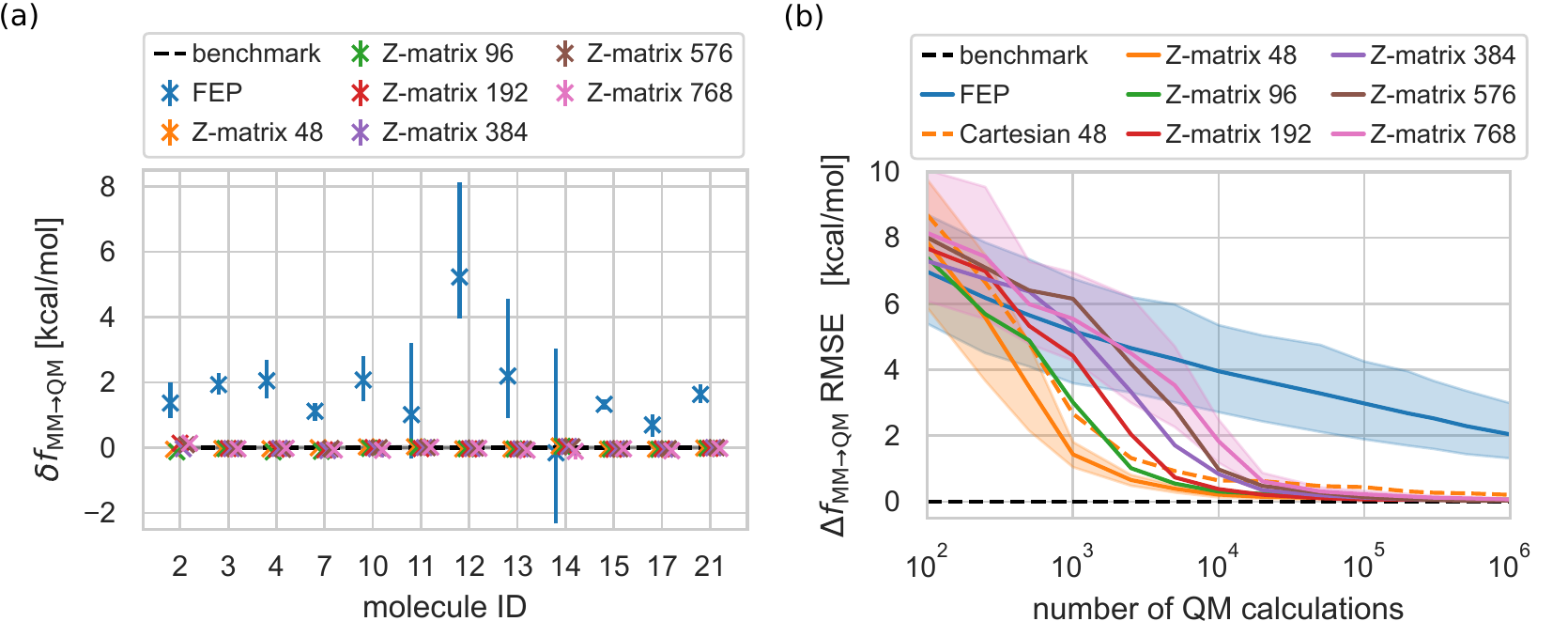}
    \centering
    \caption{\label{fig:goodset}Free energy predictions on the ``good'' HiPen subset.
    (a) Deviation of the free energy difference from the benchmark results in~\cite{scholler2022optimizing} for standard FEP and multimap TFEP for different batch sizes.
    Error bars represent bootstrap 95\% confidence intervals.
    (b) RMSD between the TFEP and reference free energy predictions for all molecules as a function of the number of single-point calculations.
    For TFEP, the NN map was trained on either Cartesian (dashed) or Z-matrix (solid) coordinates.
    For clarity, bootstrap 95\% confidence intervals are represented only for FEP and TFEP using Z-matrix coordinates and batch size 48 or 768.}
\end{figure*}

\subsection{One-epoch learning}

The multimap TFEP estimator does not remove the problem of overfitting.
Indeed, because we are generating the maps with batch training, $\map$ can converge to a function that biases the free energy estimate~\cite{rizzi2021targeted}.
To detect overfitting, one could in principle monitor the loss function on an independent validation set.
However, we note that evaluating the loss function in reference potential applications is typically so expensive that the cost of generating new data might be small or negligible in comparison.

Our solution is thus to train the NN in a single epoch.
Because each sample is seen only once during training, systematic errors due to overfitting are avoided.
If the free energy estimate is not converged once the data are exhausted, the simulations can simply be extended and the training resumed on the new data.
The pseudocode of the algorithm proposed here is illustrated in Algorithm~\ref{alg:mtfep}.
This one-epoch learning policy is not specific to the application in this work, but it might prove generally useful to avoid overfitting in other NN applications in which sampling is inexpensive compared to evaluating the validation loss, such as in Boltzmann generators~\cite{noe2019boltzmann}.

\section{Results}

We validated our method on the HiPen dataset, which was developed by Kearns \textit{et al.}~\cite{kearns2019good} specifically to test free energy calculations between different Hamiltonians.
The set includes the 22 drug-like molecules and fragments in Fig.~\ref{fig:hipen-dataset}, which are divided into a ``good"\footnote{Compound \textbf{21}, which was originally classified as ``bad", has been recently promoted to ``good" after an adjustment in the nonequilibrium protocol which made possible converging its free energy difference~\cite{scholler2022optimizing}.} (12 compounds), a ``bad" (7 compounds), and an ``ugly" (3 compounds) subset based on the difficulty observed by the authors in converging the free energy estimates for the molecules in vacuum.
For the ``good" set, the same authors provided converged free energy differences~\cite{scholler2022optimizing} between the CGenFF force field~\cite{vanommeslaeghe2010charmm} and the SCC-DFTB3/3\textit{ob} semi-empirical potential~\cite{elstner1998self,gaus2013parametrization,gaus2014parameterization,kubillus2015parameterization}, which were calculated using nonequilibrium approaches based on the Jarzynski identity (JAR)~\cite{jarzynski1997nonequilibrium} and the Crooks fluctuation theorem (CRO)~\cite{crooks2000path}.

\begin{figure*}[!ht]
    \includegraphics[width=0.9\textwidth]{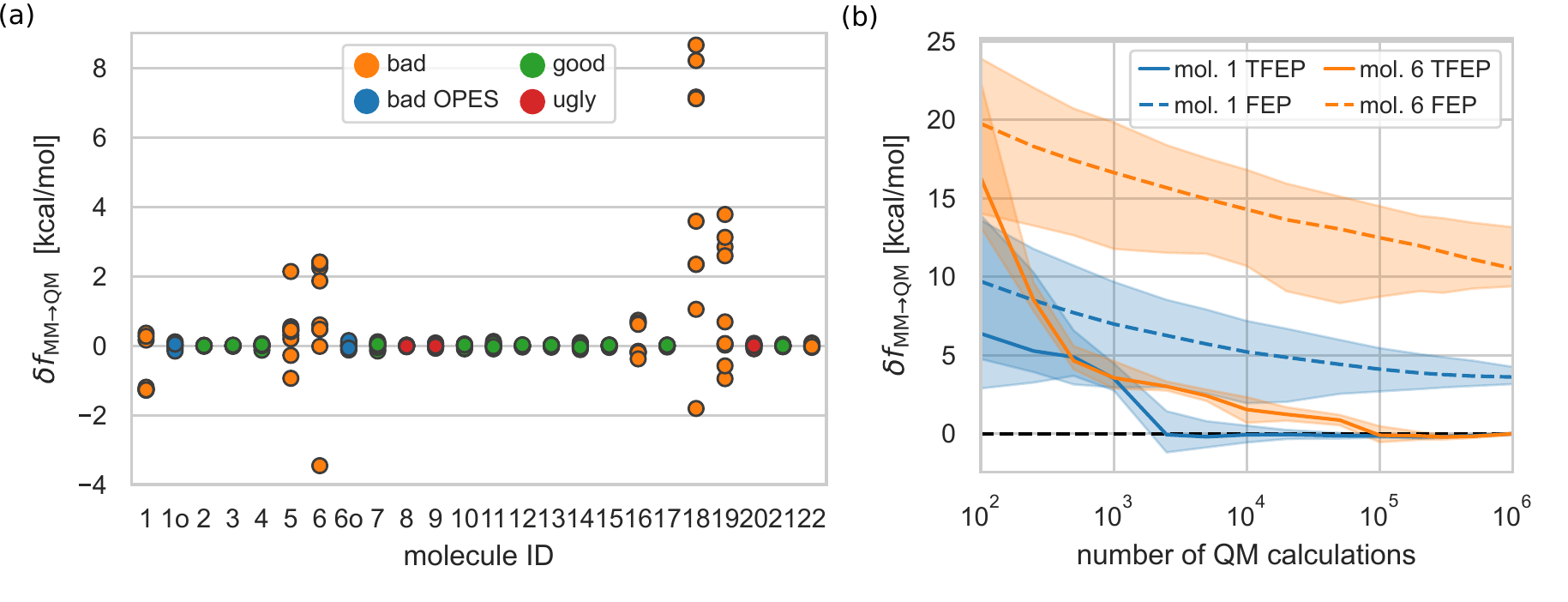}
    \centering
    \caption{\label{fig:opes-df}(a) Free energy predictions computed from the samples of the 10 independent repeats (good, bad, and ugly) or from 10 blocks for the OPES simulation (bad OPES).
    The free energies are represented as the deviation from their value computed using all the $10^6$ samples.
    For molecules \textbf{1} and \textbf{6}, we represent the deviation from the final estimate computed with OPES also for the unbiased repeats.
    (b) Free energy deviations of the multimap TFEP (solid) and FEP (dashed) estimators using OPES for molecules \textbf{1} (blue) and \textbf{6} (orange).}
\end{figure*}

\subsection{Multimap TFEP can overcome poor overlap}

We applied our methodology to the ``good" subset and compared the results to the converged CRO estimates calculated in~\cite{scholler2022optimizing}.
For the reference simulations, we followed the protocol described in~\cite{scholler2022optimizing}.
All technical details can be found in the Detailed Methods section of the Supporting Information.
Briefly, for each molecule, we ran 10 independent Langevin dynamics simulations of \SI{10}{\nano\second} in vacuum starting from 10 different configurations that we generated by randomizing the dihedral angles represented in Fig.~\ref{fig:hipen-dataset}.
We then trained an autoregressive normalizing flow NN that operated either on Cartesian or Z-matrix coordinates~\cite{noe2019boltzmann,kohler2021smooth}.
To test the robustness of our methodology, we repeated the calculation 6 times, training the normalizing flows using different batch sizes and shuffling the order of the data points.

Fig.~\ref{fig:goodset}a shows the error in the free energy differences obtained by the standard FEP and multimap TFEP estimators using Z-matrix coordinates after 1 million samples.
Standard FEP failed to converge to the benchmark results in all cases with a root mean square error (RMSE) of 2.0 [1.3, 3.0]~\si{\kcal\per\mol} (square brackets represent 95\%-percentile bootstrap confidence intervals).
In contrast, the multimap TFEP estimates agree very well with the independent nonequilibrium calculations previously published.
Regardless of the batch size, the final RMSE using Z-matrix coordinates is always below $\SI{0.1}{\kcal\per\mol}$ (see Table~\ref{tab:rmse}) and on the same order of magnitude as the uncertainties reported for the reference CRO estimates ($\sim$\SI{0.03}{\kcal\per\mol})~\cite{scholler2022optimizing}.
Neural networks operating in Cartesian coordinates also converged to the correct free energy, albeit with slightly greater RMSEs (see also Fig.~\ref{fig:si:cartesian-vs-zmatrix}).

\begin{table}[!b]
\centering
\begin{tabular}{| l | c c |}
 \hline
 Batch size & Cartesian & Z-matrix \\
 \hline\hline
 48 & 0.20 [0.11, 0.30] & 0.03 [0.03, 0.04] \\
 96 & 0.19 [0.09, 0.29] & 0.06 [0.03, 0.09] \\
 192 & 0.14 [0.06, 0.25] & 0.05 [0.03, 0.07] \\
 384 & 0.16 [0.09, 0.24] & 0.03 [0.02, 0.05] \\
 576 & 0.12 [0.07, 0.17] & 0.03 [0.02, 0.05] \\
 768 & 0.16 [0.08, 0.26] & 0.07 [0.04, 0.09] \\
 \hline
\end{tabular}
\caption{Root mean square error in \si{\kcal\per\mol} of the multimap TFEP estimates with respect to the benchmark results for the ``good" set after one million samples.
Square brackets represent 95\%-percentile bootstrap confidence intervals.}
\label{tab:rmse}
\end{table}

\subsection{Convergence is quicker with small batches and internal coordinates}

The effect of batch size and coordinate system on the convergence rate emerges clearly from inspecting the RMSE trajectories as a function of the number of samples in Fig.~\ref{fig:goodset}b.
On the ``good" set, a smaller batch size systematically resulted in faster convergence.
With batch size 48, the RMSE error after 10000 samples was reduced to 0.21 [0.13, 0.29]~\si{\kcal\per\mol}.
This initial boost for small batches is likely due to the greater number of (albeit noisier) gradient descent steps given the same number of samples.
On the other hand, in the limit of a large number of samples, smaller batch sizes did not obtain significantly smaller errors, while being less parallelizable.
In practice, a scheme starting with a small batch size that is progressively increased during the calculation might provide the best trade-off between data efficiency and parallelization.

The use of Z-matrix coordinates consistently resulted in more rapid convergence than Cartesian coordinates (see Fig.~\ref{fig:si:cartesian-vs-zmatrix}) at the price of a small computational overhead due to the conversion of the coordinates.
With 48 CPUs and a batch size of 48, this operation led to an increase of the wall-clock time by about 9\%, but this percentage would decrease if a more expensive level of theory were chosen for the target Hamiltonian.
The reason behind the higher efficiency of Z-matrix coordinates is likely their higher degree of decoupling with respect to the potential energy, which might simplify learning the task for normalizing flows.

\begin{figure*}[!h]
    \includegraphics[width=0.9\textwidth]{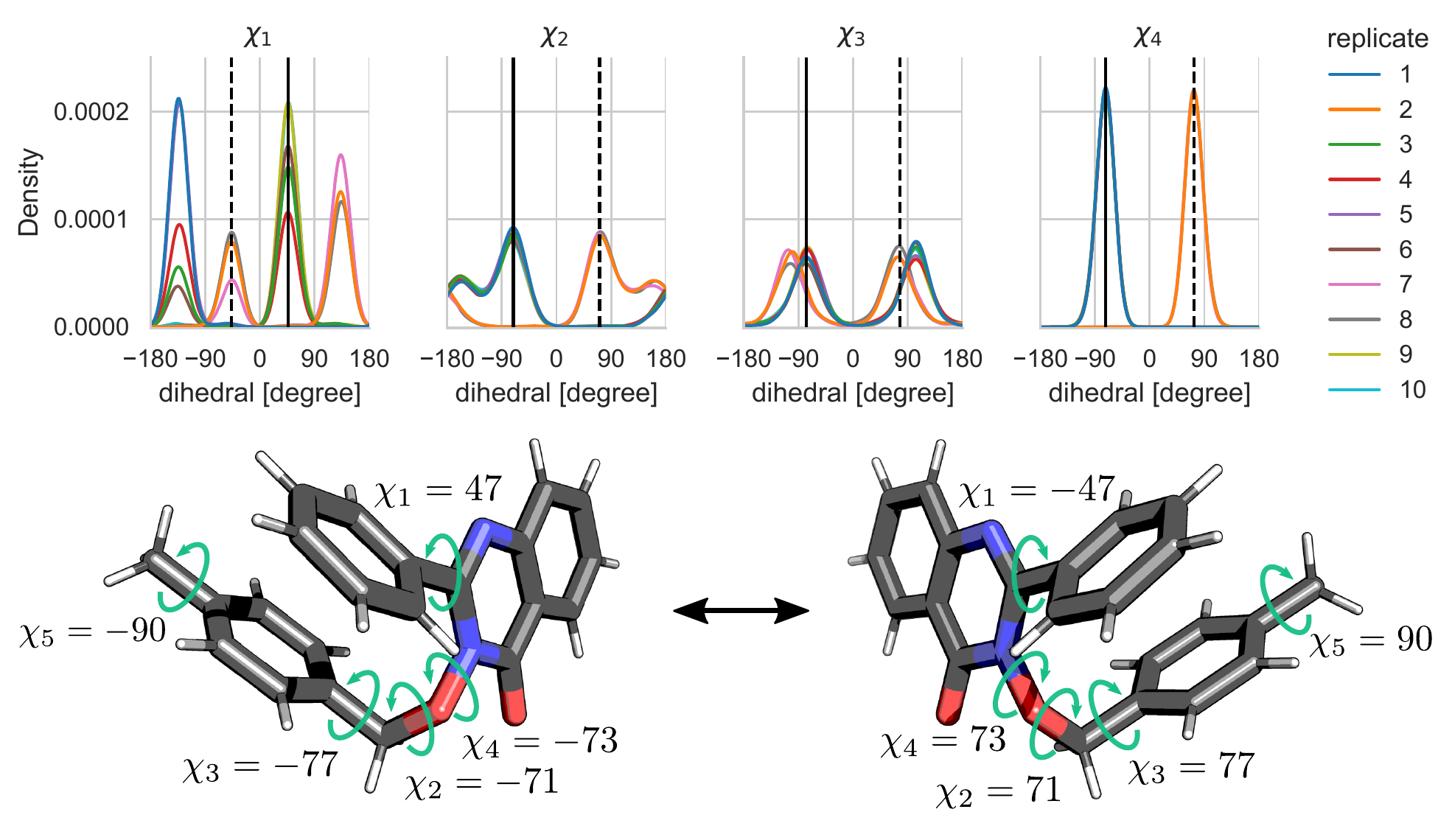}
    \centering
    \caption{\label{fig:symmetric-configurations}(Top) Distribution of the dihedrals $\chi_1$ - $\chi_4$ obtained by the 10 independent replicate simulations of molecule \textbf{14}.
    (Bottom) 3D structures of molecule \textbf{14} representing the dihedrals for two symmetric states with identical potential energy that did not interconvert during the simulations.
    The solid and dashed vertical lines in the dihedral distributions represent the values of the dihedrals in the configurations in the bottom left and right respectively.
    Note that only dihedrals $\chi_1$ - $\chi_3$ were randomized when generating the initial conditions of the replicates.}
\end{figure*}

\subsection{Multimap TFEP is cost-effective}

Compared to standard FEP, our method reached the same RMSE (\SI{2.0}{\kcal\per\mol}) with three orders of magnitude fewer samples.
Moreover, the wall-clock time per sample in our method is only marginally higher.
This is because the evaluation of the DFTB3 potential energy is the computational bottleneck of both standard FEP and multimap TFEP.
On top of this, our method must evaluate the NN (forward and backward), compute the DFTB3 forces (which enter the expression of the NN gradients), and optionally perform the conversion between Cartesian and internal coordinates.
In our experiments, however, we found this overhead to be small.
Using 48 CPUs and a batch size of 48, the overhead added on average 24\% of the wall-clock time using Cartesian coordinates and 33\% using Z-matrix coordinates.
Again, this will be dependent on the level of the target theory.

Our results also suggest that multimap TFEP is more efficient than BAR and nonequilibrium methods on this set of molecules.
The BAR calculations reported in~\cite{kearns2019good} achieved an RMSE of \SI{0.36}{\kcal\per\mol} after 1 million single-point DFTB3 calculations that must be added to the cost of the 10.01 million steps of DFTB3 molecular dynamics to generate samples from the target distribution.
Furthermore, the cheapest nonequilibrium protocol in~\cite{scholler2022optimizing} that obtained an RMSE below \SI{0.1}{\kcal\per\mol} on the ``good" set used 200 independent nonequilibrium switching trajectories of \SI{2}{\pico\second}, for a total of 400,000 DFTB3 force evaluations.
In comparison, multimap TFEP using a batch size of 192 (thus with similar parallelization capabilities) obtained the same RMSE in $\sim$50,000 force evaluations, reducing the computational requirements by a factor of 8.

\subsection{The bad set requires enhanced sampling}

We applied our method also to the ``bad" and ``ugly" subsets to investigate the origin of their difficulty.
For all these molecules, we noticed that with the original protocol~\cite{kearns2019good}, sampling was insufficient.
Indeed, contrarily to the ``good" set, the average potential energies of the 10 repeats starting from randomized configurations were always significantly different (Fig.~\ref{fig:si:avg-potential-repeats}), and this was due to poorly interconverting torsional states  (Fig.~\ref{fig:si:dihedrals-bad}) which prevented sampling convergence.

First, we should note that a few of the molecules in the ``good" subset also displayed slow interconversions between some torsional states (e.g., molecules \textbf{10}, \textbf{14}, \textbf{17}, see Fig.~\ref{fig:si:dihedrals-good}).
However, the critical difference compared to the ``bad"/``ugly" compounds is that these torsions define symmetric states with identical free energy.
An illustrative example is described in Fig.~\ref{fig:symmetric-configurations} for molecule \textbf{14}, for which the independent reference simulations could explore only one of the two symmetric states.
As we show in Appendix~\ref{sec:si:symmetry}, under easily satisfied conditions on the map, the multimap TFEP estimator in Eq.~\eqref{eq:mtfep-estimator} is invariant to the number of symmetric states that are actually sampled and their populations.
Thus, the calculated free energy differences for the ``good" compounds converged to the correct value.
Instead, the slowly-interconverting torsional states in the ``bad" and ``ugly" compounds are asymmetric, and their populations, which are effectively determined by the initial dihedral randomization of the different repeats, affect the free energy estimates.
For molecules \textbf{1}, \textbf{6}, \textbf{8}, \textbf{9}, \textbf{20}, and \textbf{22}, this problem was partially due to the accidental isomerization of a double bond during the randomization step that could not be reverted during the MD simulations.
Changing the original protocol to restrain the chirality of these double bonds solved the issue for the entire ``ugly" subset (molecules \textbf{8}, \textbf{9}, and \textbf{20}) as well as molecule \textbf{22}, and the free energies in Fig.~\ref{fig:opes-df}a reflect this updated protocol.
However, the remaining ``bad" compounds continued to display poor convergence.

These results suggest that the problem with the ``bad" subset of molecules is firstly one of sampling rather than one of poor overlap and that combining perturbation methods with enhanced sampling techniques might offer a solution.
To explore this strategy, we ran OPES~\cite{invernizzi2020rethinking} reference simulations for molecules \textbf{1} and \textbf{6} and used reweighting~\cite{bonomi2009reconstructing,invernizzi2020rethinking} to compute the averages in the loss function and the multimap free energy estimate (see also Appendix~\ref{sec:si:OPES}).
In both cases, we used a 2-dimensional collective variable, biasing dihedrals $\chi_3$ and $\chi_4$ for molecule \textbf{1} and dihedrals $\chi_1$ and $\chi_3$ for molecule \textbf{6}.
These dihedrals were chosen because they define torsional states that could not interconvert during the unbiased simulations (Fig.~\ref{fig:si:dihedrals-bad}).
We first verified that the sampling was efficient by running ten short (10~ns) OPES simulations starting from the same randomized configurations used for the unbiased dynamics.
In all the OPES different runs, the average potential energy (Fig.~\ref{fig:si:avg-potential-repeats}) and the dihedral distributions (Fig.~\ref{fig:si:dihedrals-opes}) were found to be in agreement.
We then applied our method using the data from a single longer (100~ns) OPES calculation.

Because converged free energy calculations for these molecules are unavailable, we evaluated the convergence of both sampling and $\Delta \hat{f}$ by assessing the agreement of the 10 independent free energy estimates computed for each repeat (each of $10^5$ samples) with the estimate computed using all data ($10^6$ samples).
For the OPES simulations, we instead obtained 10 estimates by dividing the 100~ns trajectory in 10 blocks of 10~ns.
Fig.~\ref{fig:opes-df}a shows that for all molecules in which sampling was effective, the estimates of the different repeats converged to the same value computed with 10 times more data.
In particular, enhanced sampling was fundamental for molecule \textbf{1} and \textbf{6}, and the free energy difference computed with OPES converged in about 2000 and $10^5$ samples respectively (see Fig.~\ref{fig:opes-df}b).

\section{Conclusions}

In this work, we have developed a method for computing free energy differences between two Hamiltonians in the presence of poor overlap between the two distributions.
The method relies on a novel estimator that generalizes TFEP to use multiple configurational maps.
Operationally, our proposed protocol is similar to standard FEP with an additional step in which the maps (and the convergence speed) are gradually improved by exploiting the information within forces.
We validated and tested the efficiency of the methodology on the HiPen dataset of small molecules by computing the free energy difference of switching from a general small molecule force field to a DFTB3 description of the potential in vacuum.
Multimap TFEP was more efficient than standard FEP and previously published nonequilibrium approaches.
This task represents a fundamental step towards obtaining binding free energies using QM(/MM) potentials, and these results open the door to successful applications in drug screening campaigns.
Moreover, we showed that combining multimap TFEP and enhanced sampling techniques can be used to overcome the problem of sampling slow degrees of freedom.
Although the number of torsional states accessible to a ligand might be limited in a water or protein environment, the problem of conformational flexibility (e.g., of small molecule drugs or protein side chains) is pervasive in binding free energy calculations, often hampering the sampling convergence~\cite{ross2020enhancing,kaus2015deal,wang2012achieving,mobley2007confine}.
Finally, the one-epoch learning policy employed here could be generally useful to avoid overfitting as an alternative to monitoring the validation loss in any machine learning setting in which generating new data is cheap compared to evaluating the loss function (e.g., in Boltzmann generators~\cite{noe2019boltzmann}) or to support on-the-fly learning strategies in other normalizing flow-based sampling schemes~\cite{invernizzi2022skipping,gabrie2022adaptive}.

The use of neural networks does not introduce any approximation, and the method converges exactly to the free energy predicted by the chosen target level of theory (and consequently also reflects its inaccuracies).
The performance of the protocol employed in this work can be improved by varying the enhanced sampling methodology~\cite{henin2022enhanced}, the learning protocol, and the architecture of the normalizing flow~\cite{papamakarios2021normalizing}.
Our tests suggest that, for small molecules, performing the normalizing flow transformation in Z-matrix coordinates works better than in Cartesian coordinates, and that the convergence of the free energy generally improves with smaller batches.
Gradually increasing the batch size during the calculation~\cite{smith2017don} might strike an optimal balance between convergence speed and parallelizability.
The latter adds to the parallelism already offered by modern molecular simulation engines, and it represents a fundamental advantage of the method given the diffusion of massively parallel computing architectures~\cite{schneider2022exascale}.
Finally, multimap TFEP can be used in conjunction with methods such as force-matching~\cite{hudson2018force,pan2019accelerated} and machine learning (reference and/or target) potentials~\cite{shen2016multiscale,chehaibou2019computing,bucko2020ab,giese2022multireference} to reduce the distance between the two distributions and obtain a cheaper model of the QM level of theory.
These solutions could significantly accelerate convergence in case the NN map were unable during training to ``discover" (and thus correct for) metastable states that are not predicted by the FF but are relevant to the QM.

The methodology can be straightforwardly applied to perturb protein-ligand systems to a QM/MM Hamiltonian, possibly including part of the protein in the QM region and mapping only a subset of the degrees of freedom (e.g., the ligand and the protein atoms close to the binding site).
Furthermore, it can be extended to different, important settings: As shown in our previous work~\cite{rizzi2021targeted}, targeted estimators can be used to predict full free energy surfaces (FES) as a function of a collective variable rather than simple free energy differences.
Thus, the method could be employed, for instance, to obtain free energy landscapes characterizing enzymatic and chemical reactions.
In general, multimap TFEP can be applied to any two levels of theory, and because the overhead due to training the configurational map is small, we expect it to be more efficient than standard FEP in all cases in which the overlap between the two Hamiltonians is poor.
Finally, as we showed in Appendix~\ref{sec:si:mtmbar}, this multimap approach can be extended to estimators such as MBAR, which is suited to cases where sampling from two or more states is equally inexpensive such as in forcefield-based alchemical free energy calculations and in generalized weighted thermodynamic perturbation~\cite{giese2022multireference}.

Given the wide scope of possible applications, ranging from materials science to drug binding, to chemical/enzymatic reactions, we believe this methodology will be useful to the community and provide a significant step towards the calculation of accurate free energies at the QM(/MM) level, as well as in other applications of NNs.

\section*{Data and code availability}

All the input files necessary to reproduce the work are available at \url{https://github.com/andrrizzi/paper-multimap-tfep-2023}.
MD and OPES trajectories, potential energies, and neural network models generated for the multimap TFEP analysis are deposited on Zenodo~\cite{rizzi2023multimap}.

\section*{Acknowledgments}
The authors gratefully acknowledge the Gauss Centre for Supercomputing e.V. (\url{www.gauss-centre.eu}) for funding this project by providing computing time through the John von Neumann Institute for Computing (NIC) on the GCS Supercomputer JUWELS~\cite{alvarez2021juwels} at J\"{u}lich Supercomputing Centre (JSC).
The project received funding from the Helmholtz European Partnering program (``Innovative high-performance computing approaches for molecular neuromedicine").
Open access publication fee funded by the Deutsche Forschungsgemeinschaft (DFG, German Research Foundation) - 491111487.
A.R. thanks Andreas Kr\"{a}mer and Michele Invernizzi for useful discussions.


\printbibliography


\clearpage

\onecolumn

\renewcommand{\theequation}{S\arabic{equation}}
\renewcommand\thepage{S\arabic{page}}
\renewcommand{\thesection}{\Alph{section}}
\renewcommand{\thetable}{S\arabic{table}}
\renewcommand{\thefigure}{S\arabic{figure}}

\setcounter{equation}{0}
\setcounter{page}{1}
\setcounter{section}{0}
\setcounter{table}{0}
\setcounter{figure}{0}

\section*{SUPPORTING INFORMATION}

\section{Multimap targeted MBAR}\label{sec:si:mtmbar}

In this section, we derive a multimap version of the MBAR estimator~\cite{shirts2008statistically,paliwal2013multistate}, which can be used when samples from $K > 1$ states are available.
The multimap version of BAR~\cite{bennett1976efficient,hahn2009using} can be seen as a special case of this estimator for $K = 2$.

We first derive the targeted MBAR estimator mainly following the derivations in~\cite{paliwal2013multistate,meng2002warp}.
Let $\x_{ki}$ be the $i$-th configuration sampled from the $k$-th state with Boltzmann weight $q_k(\x) = e^{-u_k(\x)}$ and configurational partition function $Z_k = e^{-f_k}$.
We define a set of $K$ invertible maps $\{ \map_{kr} : \Gamma_i \to \Gamma_r \}$ mapping any state $k$ into a reference state $r$.
The reference state can be arbitrary (possibly different from the $K$ sampled states) as the final equations will not depend on $r$.
From these maps, we can construct $K \times K$ functions mapping any two sampled states $i,j$
\begin{equation}
    \map_{ij}(\x) = \map_{rj}(\map_{ir}(\x)) \; ,
\label{eq:si:tmbar-maps}
\end{equation}
where we defined $\map_{rj} = \map^{-1}_{jr}$.
We indicate the Jacobian determinant of $\map_{ij}(\x)$ with $|J_{ij}(\x)|$.

Let $\alpha(\x)$ be any strictly positive function defined on $\Gamma_r$.
Then, the following identity holds
\begin{align}\label{eq:si:tmbar-basic-identity}
\begin{split}
    \MoveEqLeft[5]  Z_i \mean{ q_j(\map_{ij}(\x)) |J_{rj}(\map_{ir}(\x))| \alpha_j(\map_{ir}(\x)) }_i \\
    ={}& \int_{\Gamma_i} q_i(\x) q_j(\map_{ij}(\x)) |J_{rj}(\map_{ir}(\x))| \alpha_j(\map_{ir}(\x)) d\x \\
    ={}& \int_{\Gamma_j} q_i(\map_{ji}(\y)) q_j(\y) |J_{ri}(\map_{jr}(\y))| \alpha_j(\map_{jr}(\y)) d\y \\
    ={}& Z_j \mean{q_i(\map_{ji}(\y)) |J_{ri}(\map_{jr}(\y))| \alpha_j(\map_{jr}(\y)) }_j
\end{split}
\end{align}
where in the third line we applied a change of variable $\x = \map_{ji}(\y)$ and we used
\begin{equation}
    |J_{rj}(\map_{ir}(\map_{ji}(\y)))| |J_{ji}(\y)| = |J_{ri}(\map_{jr}(\y))| \; .
\end{equation}
By summing both sides over index $j$ and estimating the mean with a sample average, one obtains an equation for the estimates of $\hat{Z}_i$
\begin{multline}
    \sum_{j=1}^K \frac{\hat{Z}_i}{N_i} \sum_{n=1}^{N_i} q_j(\map_{ij}(\x_{in})) |J_{rj}(\map_{ir}(\x_{in}))| \alpha_j(\map_{ir}(\x_{in})) \\
    = \sum_{j=1}^K \frac{\hat{Z}_j}{N_j} \sum_{n=1}^{N_j}  q_i(\map_{ji}(\x_{jn})) |J_{ri}(\map_{jr}(\x_{jn}))| \alpha_j(\map_{jr}(\x_{jn})) \; .
\label{eq:si:tmbar-equation-with-alpha}
\end{multline}
The function $\alpha_j$ is typically chosen to minimize the variance of the estimator~\cite{shirts2008statistically}.
Here we choose
\begin{equation}
    \alpha_j(\x) = \frac{N_j \hat{Z}^{-1}_j}{\sum_{k=1}^K N_k Z^{-1}_k q_k(\map_{rk}(\x)) |J_{rk}(\x)|}
\label{eq:si:alpha}
\end{equation}
Plugging the expression of $\alpha_j(\x)$ in Eq.~\eqref{eq:si:tmbar-equation-with-alpha}, we obtain
\begin{multline}
    \sum_{j=1}^K\frac{\hat{Z}_i}{N_i} \sum_{n=1}^{N_i} \frac{q_j(\map_{ij}(\x_{in})) |J_{rj}(\map_{ir}(\x_{in}))| N_j \hat{Z}^{-1}_j}{\sum_{k=1}^K N_k Z^{-1}_k q_k(\map_{ik}(\x_{in})) |J_{rk}(\map_{ir}(\x_{in}))|} \\
    = \sum_{j=1}^K \frac{\hat{Z}_j}{N_j} \sum_{n=1}^{N_j}  \frac{q_i(\map_{ji}(\x_{jn})) |J_{ri}(\map_{jr}(\x_{jn}))| N_j \hat{Z}^{-1}_j}{\sum_{k=1}^K N_k Z^{-1}_k q_k(\map_{jk}(\x_{jn})) |J_{rk}(\map_{jr}(\x_{jn}))|} \; .
\label{eq:si:tmbar-equation}
\end{multline}
By switching the summation order on the left side and multiplying and dividing the summands of the right hand by $|J_{jr}(\x_{jn})|$, we obtain $K$ equations
\begin{equation}
    \hat{Z}_i = \sum_{j=1}^K \sum_{n=1}^{N_j}  \frac{q_i(\map_{ji}(\x_{jn})) |J_{ji}(\x_{jn})|}{\sum_{k=1}^K N_k Z^{-1}_k q_k(\map_{jk}(\x_{jn})) |J_{jk}(\x_{jn})|} \; .
\label{eq:si:tmbar}
\end{equation}
which are usually solved for $\hat{Z}_i$ in a self-consistent fashion~\cite{shirts2008statistically}.
Eq.~\eqref{eq:si:tmbar} corresponds to the estimator derived in~\cite{paliwal2013multistate}.
Note that, while Eq.~\eqref{eq:si:tmbar} does not depend on the reference state $r$, the way we constructed these maps comes with some implicit assumption.
For example, that $\map_{ij} = \map_{ji}^{-1}$ for any pair $i,j$.

We now extend Eq.~\eqref{eq:si:tmbar} to the multimap case by noting that Eq.~\eqref{eq:si:tmbar-basic-identity} is true for any set of maps defined above.
We can thus generate $B$ sets of maps $\{ \map^b_{kr} \}$ and sum both sides of Eq.~\eqref{eq:si:tmbar-basic-identity} over $b$
\begin{align}\label{eq:si:mtmbar-basic-identity}
\begin{split}
    \MoveEqLeft[5]  Z_i \sum_{b=1}^B \mean{ q_j(\map^b_{ij}(\x)) |J^b_{rj}(\map^b_{ir}(\x))| \alpha^b_j(\map^b_{ir}(\x)) }_i \\
    ={}& Z_j \sum_{b=1}^B \mean{q_i(\map^b_{ji}(\y)) |J^b_{ri}(\map^b_{jr}(\y))| \alpha^b_j(\map^b_{jr}(\y)) }_j \; ,
\end{split}
\end{align}
where we have used $\alpha^b_j(\x)$ to represent the same function in Eq.~\eqref{eq:si:alpha} but using the $\map^b$ maps.
We can go through the same steps in Eq.~\eqref{eq:si:tmbar-equation-with-alpha}-\eqref{eq:si:tmbar} to obtain its multimap version
\begin{equation}
    \hat{Z}_i = \frac{1}{B} \sum_{b=1}^B \sum_{j=1}^K \sum_{n=1}^{N_j}  \frac{q_i(\map^b_{ji}(\x_{jn})) |J^b_{ji}(\x_{jn})|}{\sum_{k=1}^K N_k Z^{-1}_k q_k(\map^b_{jk}(\x_{jn})) |J^b_{jk}(\x_{jn})|} \; .
\label{eq:si:mtmbar}
\end{equation}

\section{Multimap TFEP does not require sampling all symmetric states}\label{sec:si:symmetry}

First, we define a symmetric transformation an invertible bijection $f: \Gamma \to \Gamma$ that is mass-preserving (i.e., the absolute value of its Jacobian determinant is 1) and leaves invariant the potential energy of the system
\begin{equation}\label{eq:si:symmetric-operation}
	u(\x) = u(f(\x))
\end{equation}
for any configuration $\x \in \Gamma$.
As an example, if all bonds and valence angles of molecule \textbf{17} were constrained to a fixed value, a simple rotation by 180 degrees of either dihedral $\chi_1$ or $\chi_2$ would be a symmetric transformation.
With flexible bonds and angles, the transformation would also need to shift the atom positions in order to obtain an identical configuration (up to a permutation of atoms).
For future reference, we call these transformation $f^{(\chi_1)}$ and $f^{(\chi_2)}$.

We say that a system has $M$ symmetric states $\Gamma^{(m)} \subset \Gamma$, $m=1,\dots,M$ with respect to a set of symmetric transformations $F = \{ f^{(\alpha)} \}$ if:
\begin{enumerate}
	\item The set of all symmetric states $G = \{ \Gamma^{(m)} \}$ form a partition of $\Gamma$.
	\item For any pair $\Gamma^{(i)}, \Gamma^{(j)}$ there is a symmetric operation $f \in F$
\begin{equation}\label{eq:si:symmetric-state-condition1}
	\Gamma^{(j)} = \{ f(\x) | \x \in \Gamma^{(i)} \} \; .
\end{equation}
\end{enumerate}

Note that there is no unique definition of symmetric states for a particular system.
First, the symmetric states depend on the set of symmetric transformations.
Second, even with the same set of symmetric transformations, there can be multiple partitions satisfying 1-2.
In the example of molecule \textbf{17}, we can choose 4 transformations: the identity, $f^{(\chi_1)}$, $f^{(\chi_2)}$, and their composition $f^{(\chi_1)} \circ f^{(\chi_2)}$.
These give rise to 4 symmetric states, which we can generate by applying the transformations in $F$ to a single symmetric state including all configurations such that $\chi_1$ and $\chi_2$ are within $[\psi, \psi+180)$ for any choice of $\psi$.


Consider now two potential energy functions $u_1(\x)$ and $u_2(\x)$ with the same symmetries.
Assume also that the TFEP map $\map$ transforms the chosen symmetric states into other symmetric states.
I.e., the set $G_{\map} = \{ \Gamma_{\map}^{(m)} \}$ obtained by applying $\map$ to each symmetric state $\Gamma^{(i)}$
\begin{equation}\label{eq:si:transformed-symmetric-state}
	\Gamma_{\map}^{(i)} = \{ \map(\x) | \x \in \Gamma^{(i)} \}
\end{equation}
is also a set of (possibly different) symmetric states satisfying conditions 1-2.
We show here that under these assumptions the (multimap) TFEP estimator  is invariant to the number of symmetric states that are effectively sampled.

First, because the symmetric states form a partition of $\Gamma$ we have
\begin{equation}\label{eq:si:partition-function-decomposition}
	Z\starget = \sum_{i=1}^M \int_{\Gamma^{(i)}} e^{-u\starget(\x)} d\x \; .
\end{equation}
Second, using condition 2, we can show that the free energy of each symmetric state is identical since
\begin{equation}\label{eq:si:symmetric-state-identical-free-energy}
\begin{split}
	\int_{\Gamma^{(j)}} e^{-u\starget(\y)} d\y &= \int_{\Gamma^{(i)}} e^{-u\starget(f(\x))} d\x \\
	&= \int_{\Gamma^{(i)}} e^{-u\starget(\x)} d\x \; ,
\end{split}
\end{equation}
where in the first line we used that the Jacobian determinant of $f$ is 1 and in the second line we used Eq.~\eqref{eq:si:symmetric-operation}.
Thus, using Eq.~\eqref{eq:si:partition-function-decomposition} and Eq.~\eqref{eq:si:symmetric-state-identical-free-energy} together, we obtain the configurational partition function of each symmetric state
\begin{equation}\label{eq:si:symmetric-state-partition-function}
	\int_{\Gamma^{(i)}} e^{-u\starget(\x)} d\x = \frac{Z\starget}{M} \; .
\end{equation}
Note that this result is valid regardless of how the symmetric states are chosen, and, in particular, it is valid also for the symmetric states generated by the map $\{ \Gamma_{\map}^{(m)} \}$.

Now, assume that we only have samples from a subset of $L < M$ symmetric states.
This is equivalent to sampling from a distribution
\begin{equation}\label{eq:si:prefL}
	p\sref^{(L)}(\x) = \frac{e^{-u\sref(\x)}}{Z\sref^{(L)}}
\end{equation}
which is defined on $\Gamma^{(L)} = \cup_{l=1}^L \Gamma^{(l)}$ and whose normalization constant $Z\sref^{(L)}$ is related to the configurational partition function by
\begin{equation}\label{eq:si:ZrefL}
	Z\sref^{(L)} = \int_{\Gamma^{(L)}} e^{-u\sref(\x)} d\x = \sum_{l=1}^L \int_{\Gamma^{(l)}} e^{-u\sref(\x)} d\x = \frac{L Z\sref}{M} \; .
\end{equation}
By applying TFEP to this distribution we obtain
\begin{equation}\label{eq:si:mtfep-estimator-subset}
\begin{split}
	\int_{\Gamma^{(L)}} p\sref^{(L)}(\x) e^{-\w{\x}} d\x &= \frac{M}{ LZ\sref} \sum_{l=1}^L \int_{\Gamma^{(l)}} e^{-u\starget(\map(\x)) + \logJm{\x}} d\x \\
	&= \frac{M}{L Z\sref}\sum_{l=1}^L \int_{\Gamma_{\map}^{(l)}} e^{-u\starget(\y)}d\y \\
	&= \frac{Z\starget}{Z\sref}
\end{split}
\end{equation}
where we used Eq.~\eqref{eq:si:transformed-symmetric-state} in the second line, and in the third line we used that Eq.~\eqref{eq:si:symmetric-state-partition-function} is valid also for the mapped symmetric states $G_{\map}$.
Finally, it is trivial to show that Eq.~\eqref{eq:si:mtfep-estimator-subset} is true also for the multimap TFEP estimator if all the maps satisfy the conditions above.

\section{Multimap TFEP with enhanced sampling}\label{sec:si:OPES}

We show here that the multimap TFEP estimator can be used in combination with biased simulations.
In particular, ensemble averages such as that in Eq.~\eqref{eq:tfep} can be computed in OPES simulations by applying a reweighting scheme similar to that in umbrella sampling.
Thus, using $B$ maps, an estimate of the free energy difference can be computed with
\begin{equation}\label{eq:si:opes-mtfep}
    e^{- \Delta f\sdiff} = \frac{1}{B} \sum_b^B \frac{\mean{ e^{v(\bm{s})} e^{-\w[\map^b]{\x}} }_O}{\mean{ e^{v(\bm{s})} }_O} \; .
\end{equation}
where $\mean{\cdot}_O$ represents the time average in the OPES ensemble, and $v(\bm{s})$ is the bias potential (divided by $k_B T$) as a function of a collective variable (CV) $\bm{s}$.

In our methods, the maps are generated during training, and we can estimate the average in the numerator using the samples in a batch of size $L$
\begin{equation}\label{eq:si:opes-mtfep-numerator}
    \mean{e^{v(\bm{s})} e^{-\w[\map^b]{\x}} }_O \approx \frac{1}{L} \sum_l^L e^{v(\bm{s}_{bl})} e^{-\w[\map^b]{\x_{bl}}} \; ,
\end{equation}
where $\bm{s}_{bl}$ is the CV evaluated for the $l$-th sample in the $b$-th batch.
The average in the denominator does not depend on the map, and we can estimate it using all the data.
\begin{equation}\label{eq:si:opes-mtfep-denominator}
    \mean{ e^{v(\bm{s})} }_O \approx \frac{1}{BL} \sum_b^B\sum_l^L e^{v(\bm{s}_{bl})}
\end{equation}
In practice, Eq.~\eqref{eq:si:opes-mtfep-numerator} and Eq.~\eqref{eq:si:opes-mtfep-denominator} can be evaluated in logarithmic units (e.g., using \texttt{logsumexp}) for numerical stability.
Putting these equations together we can write the estimator
\begin{equation}\label{eq:si:opes-mtfep-estimator}
    e^{- \Delta f\sdiff} \approx \sum_b^B\sum_l^L \frac{ e^{v(\bm{s}_{bl})} e^{-\w[\map^b]{\x_{bl}}} }{ \sum_b^B\sum_l^L e^{v(\bm{s}_{bl})} } \; .
\end{equation}

\section{Detailed methods}\label{sec:si:detailed-methods}

All the input files necessary to reproduce the work are available at \url{https://github.com/andrrizzi/paper-multimap-tfep-2023}.

\subsection{Reference molecular dynamics simulations}

We set up and ran the reference molecular dynamics simulations in gas phase following in part the protocol described in~\cite{scholler2022optimizing} using the CHARMM~\cite{brooks2009charmm} input files provided with the HiPen dataset~\cite{kearns2018hipen}.
We generated 10 independent starting configurations for each molecule by randomizing the dihedrals indicated as arrows in Fig.~\ref{fig:hipen-dataset}.
Each configuration was minimized in vacuum using the adopted basis Newton-Raphson method~\cite{chu2003super} for 1,000 steps while restraining the dihedrals to their randomized values using a harmonic potential with a force constant of 100~kcal~mol$^{-1}$~rad$^{-2}$.
Contrarily to the original protocol, we also randomized the $\chi_2$ dihedral of compounds \textbf{2} and \textbf{7} (see Fig.~\ref{fig:hipen-dataset}) as they were found to be slow degrees of freedom~\cite{scholler2022optimizing}.
Because this randomization protocol sometimes led to the isomerization of a double bond for several molecules in the ``bad" and ``ugly" sets, the dihedrals controlling the chirality of the double bonds were restrained to their initial value.
For each randomized configuration, we then ran 15~ns of Langevin dynamics at 300~K using a time step of \SI{1}{\femto\second} and a collision frequency of \SI{5}{\per\pico\second}.
The first 5~ns of each simulation were discarded, thus leaving a total of 100~ns of MD data for the free energy estimation.
All the unbiased reference simulations were performed using the CHARMM suite (version 46b1)~\cite{brooks2009charmm}.

For molecules \textbf{1} and \textbf{6}, we also ran an OPES~\cite{invernizzi2020rethinking} simulation using AMBER2020~\cite{case2020amber} and PLUMED~2.8.1~\cite{tribello2014plumed} after converting the CHARMM input files to AMBER format with ParmEd~\cite{shirts2017lessons}.
To validate the input files conversion, we compared the AMBER and CHARMM potential energies over 100 random configurations and found negligible deviations (on the order of $10^{-6}$~\si{\kcal\per\mol}).
For each molecule, we used two dihedrals as collective variables to bias: $\chi_3$ and $\chi_4$ for molecule \textbf{1} and $\chi_1$ and $\chi_3$ for molecule \textbf{6}, which we found to be slow degrees of freedom in the unbiased dynamics.
The energy barrier parameter of OPES was set to \SI{20}{\kcal\per\mol} and the probability density was estimated with a pace of 500 time steps.
In this case, a single simulation of 105~ns was performed, and the first 5~ns were discarded to obtain the same total amount of data as the unbiased MD.

\subsection{TFEP analysis}

The new multimap TFEP method was implemented in a freely available (\url{https://github.com/andrrizzi/tfep}) Python library (version 0.1).
The code relies on PyTorch~\cite{paszke2019pytorch} to implement and train the normalizing flow maps and on ASE~\cite{larsen2017atomic} to obtain potential energies and forces from external molecular simulation engines.

We built a dataset of configurations for each molecule by subsampling the reference MD trajectories at 0.1~ps intervals for a total of 1 million frames per molecule.
These structures provided the data for both the standard and targeted FEP estimators.
For the target potential, we used DFTB3 with 3ob-3-1 parameters to be able to validate multimap TFEP against the benchmark results in~\cite{scholler2022optimizing}.

To evaluate DFTB3/3ob potential energies and forces, we chose the AMBER implementation over the CHARMM implementation used in~\cite{scholler2022optimizing} because a Python interface for AMBER was readily available in ASE.
To verify that differences in free energy estimates would not arise due to differences in the two implementations, we also compared the AMBER and CHARMM DFTB3 potential energies over 100 random configurations for each molecule in the dataset.
A molecule-dependent but constant offset between the two potentials emerged (see Table~\ref{tab:si:amber-offset}), which we accounted for when comparing TFEP and benchmark results.

To analyze the performance of the method as a function of the number of samples $N$, we computed 95\% percentile confidence intervals for the free energy estimates with bootstrapping by resampling $N$ work values 2000 times.
For standard FEP, the resampling was always performed considering all work values available for the molecule.
For targeted FEP, however, the variance and bias of the generalized work decrease dramatically as the map is trained, and mixing samples generated at the beginning and the end of the training would result in artificially small errors.
Thus, to evaluate confidence intervals after $N$ samples, we resampled only from the work values generated during the first $N/L$ mini-batches (where $L$ is the batch size).

The program was run on the standard computing nodes of the JUWELS supercomputer Cluster module~\cite{alvarez2021juwels}, which are configured with 48 physical cores and 96~GB of RAM.
We executed the neural network forward/backward pass on a single node using PyTorch native CPU parallelization scheme, while we distributed the calculations of energies and forces across batches by launching one parallel execution of AMBER for each batch configuration.

\subsection{Normalizing flow}

As input coordinates of the normalizing flow, we tested both Cartesian and Z-matrix internal coordinates.
In the first case, the molecule was first translated and rotated using three atoms to represent all coordinates in a local frame of reference.
The removed 6 degrees of freedom were not passed as input to the normalizing flow.
In the latter case, Cartesian coordinates were first mapped to Z-matrix coordinates, passed through the flow, and then mapped back to Cartesian.
The conversion from and to internal coordinates and its Jacobian was computed using the bgflow~\cite{noe2019boltzmann} (\url{https://github.com/noegroup/bgflow}) library.

We automatically generated the Z-matrix of each molecule through a simple heuristic that avoided selecting triplets of collinear atoms, which resulted in undefined angles and instabilities during training.
Specifically, we first converted the molecule into a graph based on bond connectivity (i.e., Euclidean distances between atoms were discarded).
The first atom was picked as the graph center.
Next, the atoms were placed in the Z-matrix in the order explored by traversing the graph breadth-first.
For each atom, the bond, angle, and dihedral atoms were selected among those previously inserted in the Z-matrix according to the following priorities, in this order: 1) closest to the inserted atom; 2) only for angle and dihedral atoms, closest to the bond atom; 3) only for heavy atoms, hydrogens were de-prioritized.
In particular, 2) was necessary to avoid selecting triplets of collinear atoms, which resulted in undefined angles and instabilities during training.

The mapping function was implemented using a masked autoregressive network (MAF) architecture~\cite{papamakarios2017masked} with 6 layers, each composed of a 2-layer MADE~\cite{germain2015made} conditioner.
The width of each MADE layer was set to 16 times the number of degrees of freedom of the molecule to generate the correct number of parameters for the neural spline transformer~\cite{durkan2019neural} described below.
The order of the atoms was inverted at each layer so that the mapped coordinates depended on all inputs.
We chose to use neural spline transformers for both Z-matrix and Cartesian coordinates because of their ability to handle periodic coordinates~\cite{rezende2020normalizing} (such as angles) and to compare the two types of coordinates using transformations with the same level of expressive power.
Neural splines can parameterize invertible functions by interpolating between fixed points (or ``knots") at which the function must have a predefined slope using rational-quadratic splines.
The position of the first and last knots are normally given by the user, and they define the input/output domain of the function, while slopes and positions of all other knots are generated by the conditioner network.
All neural splines used 6 knots.
We defined the input and output domain of the neural spline transformation for each Cartesian coordinate $x_i$ as $(x_i^{\mathrm{min}} - \SI{1.5}{\angstrom}, x_i^{\mathrm{max}} + \SI{1.5}{\angstrom})$, where $x_i^{\mathrm{min}/\mathrm{max}}$ are minimum and maximum value of the coordinate in the reference MD simulation respectively.
For internal coordinates, we defined the domain of the bond length, angle, and dihedral transformations as $(\SI{0.5}{\angstrom}, \SI{3.0}{\angstrom})$, $(\SI{0.0}{\radian}, \SI{\pi}{\radian})$, and $(\SI{0.0}{\radian}, 2\SI{\pi}{\radian})$ respectively.
For angles and dihedrals, the slope of the last knot was set equal to that of the first knot to ensure smooth transformations across periodic boundaries~\cite{rezende2020normalizing}.
Furthermore, neural spline transformations of periodic coordinates were composed with an initial phase translation as suggested in~\cite{rezende2020normalizing}, which was also generated by the MADE network.

The network was initialized to represent the identity function and trained for a single epoch using the AdamW~\cite{loshchilov2018decoupled} optimizer with hyperparameters $\beta_1 = 0.9$, $\beta_2=0.999$, weight decay coefficient 0.01, and a learning step of 0.001.

\clearpage

\section{Tables}

\begin{table}[!h]
\centering
\begin{tabular}{| c c c |}
 \hline
 molecule ID & ZINC ID & offset [\si{\kcal\per\mol}] \\ [0.5ex] 
 \hline\hline
 1 & 00061095 & -25313.64 \\
 \hline
 2 & 00077329 & -13673.17 \\
 \hline
 3 & 00079729 & -15350.78 \\
 \hline
 4 & 00086442 & -17104.43 \\
 \hline
 5 & 00087557 & -21931.99 \\
 \hline
 6 & 00095858 & -22414.55 \\
 \hline
 7 & 00107550 & -14823.26 \\
 \hline
 8 & 00107778 & -19513.09 \\
 \hline
 9 & 00123162 & -25449.31 \\
 \hline
 10 & 00133435 & -23980.56 \\
 \hline
 11 & 00138607 & -25581.77 \\
 \hline
 12 & 00140610 & -14864.68 \\
 \hline
 13 & 00164361 & -17946.56 \\
 \hline
 14 & 00167648 & -29818.87 \\
 \hline
 15 & 00169358 & -19353.82 \\
 \hline
 16 & 01755198 & -16782.32 \\
 \hline
 17 & 01867000 & -18766.90 \\
 \hline
 18 & 03127671 & -29933.52 \\
 \hline
 19 & 04344392 & -34965.23 \\
 \hline
 20 & 04363792 & -25445.86 \\
 \hline
 21 & 06568023 & -21517.02 \\
 \hline
 22 & 33381936 & -26406.08 \\
 \hline
\end{tabular}
\caption{Constant offset between the AMBER and CHARMM implementation of the DFTB3/3ob potential for all the molecules in the HiPen dataset.
The offset was computed as the average difference of the potential energies computed on 100 configurations selected randomly from the reference MD simulation.
The maximum deviation from the offset value reported in the table due to numerical precision across the random samples was always below \SI{0.01}{\kcal\per\mol}.}
\label{tab:si:amber-offset}
\end{table}

\begin{table}[!h]
\centering
\begin{tabular}{| c c c c |}
\hline
molecule ID & ZINC ID & $\Delta \hat{f}$ [\si{\kcal\per\mol}] & $\sigma\Delta \hat{f}$ [\si{\kcal\per\mol}] \\ [0.5ex]
\hline\hline
1 & 00061095 & -29301.96 [-29302.02, -29301.90] & 0.082 \\
\hline
2 & 00077329 & -15257.39 [-15257.40, -15257.39] & 0.008 \\
\hline
3 & 00079729 & -17414.15 [-17414.16, -17414.14] & 0.009 \\
\hline
4 & 00086442 & -19256.33 [-19256.38, -19256.30] & 0.050 \\
\hline
6 & 00095858 & -25120.45 [-25120.52, -25120.39] & 0.085 \\
\hline
7 & 00107550 & -16992.71 [-16992.78, -16992.66] & 0.084 \\
\hline
8 & 00107778 & -21986.62 [-21986.63, -21986.61] & 0.014 \\
\hline
9 & 00123162 & -29451.81 [-29451.85, -29451.78] & 0.048 \\
\hline
10 & 00133435 & -28337.87 [-28337.89, -28337.85] & 0.045 \\
\hline
11 & 00138607 & -29461.99 [-29462.03, -29461.95] & 0.059 \\
\hline
12 & 00140610 & -17076.26 [-17076.27, -17076.25] & 0.018 \\
\hline
13 & 00164361 & -20567.67 [-20567.68, -20567.65] & 0.020 \\
\hline
14 & 00167648 & -35082.49 [-35082.55, -35082.44] & 0.066 \\
\hline
15 & 00169358 & -22407.68 [-22407.69, -22407.66] & 0.020 \\
\hline
17 & 01867000 & -22673.42 [-22673.43, -22673.40] & 0.017 \\
\hline
20 & 04363792 & -28700.59 [-28700.62, -28700.57] & 0.048 \\
\hline
21 & 06568023 & -25069.02 [-25069.04, -25069.01] & 0.020 \\
\hline
22 & 33381936 & -30178.60 [-30178.63, -30178.58] & 0.039 \\ [1ex]
\hline\hline
5 & 00087557 & -25596.41 [-25596.48, -25596.35] & 0.731 \\
\hline
16 & 01755198 & -19633.18 [-19633.23, -19633.15] & 0.441 \\
\hline
18 & 03127671 & -34525.32 [-34526.39, -34523.73] & 3.370 \\
\hline
19 & 04344392 & -40881.92 [-40882.36, -40881.47] & 1.644 \\
\hline
\end{tabular}
\caption{Final estimates for the free energy differences for all molecules computed with our multimap TFEP approach with batch size 48 and Z-matrix coordinates.
Numbers between squared brackets represent 95\%-percentile bootstrap confidence intervals.
The last column shows the standard deviations of the 10 estimates computed separately for the 10 repeats.
For molecules \textbf{1} and \textbf{6}, we report the calculations performed with the OPES reference simulations.
For completeness, we report also the estimates for molecules \textbf{5}, \textbf{16}, \textbf{18}, and \textbf{19}, although we consider them unconverged.
The free energies include the offsets reported in Table~\ref{tab:si:amber-offset} for easiness of comparison with the CHARMM results in the literature.}
\label{tab:si:df-predictions}
\end{table}

\begin{table}[!h]
\centering
\begin{tabular}{|ccccc|}
\hline
 molecule ID &  Cartesian (batch=48) &  Z-Matrix (batch=48) &  Cartesian (batch=768) &  Z-Matrix (batch=768) \\ [0.5ex]
\hline\hline
           1 &              8.359929 &             8.402400 &               4.447347 &              4.085549 \\
           \hline
           2 &              6.111121 &             6.301148 &               4.899921 &              3.627578 \\
           \hline
           3 &              6.051466 &             7.136897 &               5.233341 &              3.664553 \\
           \hline
           4 &              6.192607 &             7.081699 &               3.567053 &              3.585947 \\
           \hline
           5 &              6.414840 &            10.202600 &               3.605561 &              3.750300 \\
           \hline
           6 &              8.192296 &             8.341677 &               4.034204 &              3.942590 \\
           \hline
           7 &              6.216101 &             6.877735 &               3.558637 &              3.587464 \\
           \hline
           8 &              6.761458 &             8.107923 &               3.861094 &              3.695925 \\
           \hline
           9 &              7.179133 &             8.114956 &               3.956955 &              3.889869 \\
           \hline
          10 &              6.424598 &             7.878607 &               3.745444 &              3.775938 \\
          \hline
          11 &              7.783509 &             7.847552 &               3.791934 &              3.808900 \\
          \hline
          12 &              6.430227 &            15.561238 &               3.851895 &              3.677079 \\
          \hline
          13 &              5.636293 &             6.977579 &               3.509684 &              3.657412 \\
          \hline
          14 &              8.104997 &             8.969143 &               3.795184 &              4.405756 \\
          \hline
          15 &              6.052706 &             6.985440 &               3.671756 &              3.569883 \\
          \hline
          16 &              9.506134 &             6.551579 &               3.530345 &              3.560771 \\
          \hline
          17 &              5.412491 &             6.148449 &               3.565387 &              3.708041 \\
          \hline
          18 &             10.426524 &             8.739259 &               4.005366 &              4.726245 \\
          \hline
          19 &             14.717742 &             9.964059 &               4.075284 &              4.387286 \\
          \hline
          20 &              7.279700 &             8.316693 &               3.872957 &              3.928537 \\
          \hline
          21 &              6.219947 &             7.494313 &               3.554381 &              3.845970 \\
          \hline
          22 &             11.509201 &             9.283224 &               4.136425 &              3.875788  \\
          \hline
\end{tabular}
\caption{Wall-clock time (in hours) required to run the multimap TFEP analysis on the full dataset (1 million samples) on a single node (48 CPUs) of the JUWELS cluster for different types of input coordinates and batch sizes.
Even on a single node, larger batch sizes can take advantage of more efficient parallelization and reduced disk overhead.}
\end{table}

\clearpage

\section{Figures}

\begin{figure}[!h]
    \includegraphics[width=0.9\textwidth]{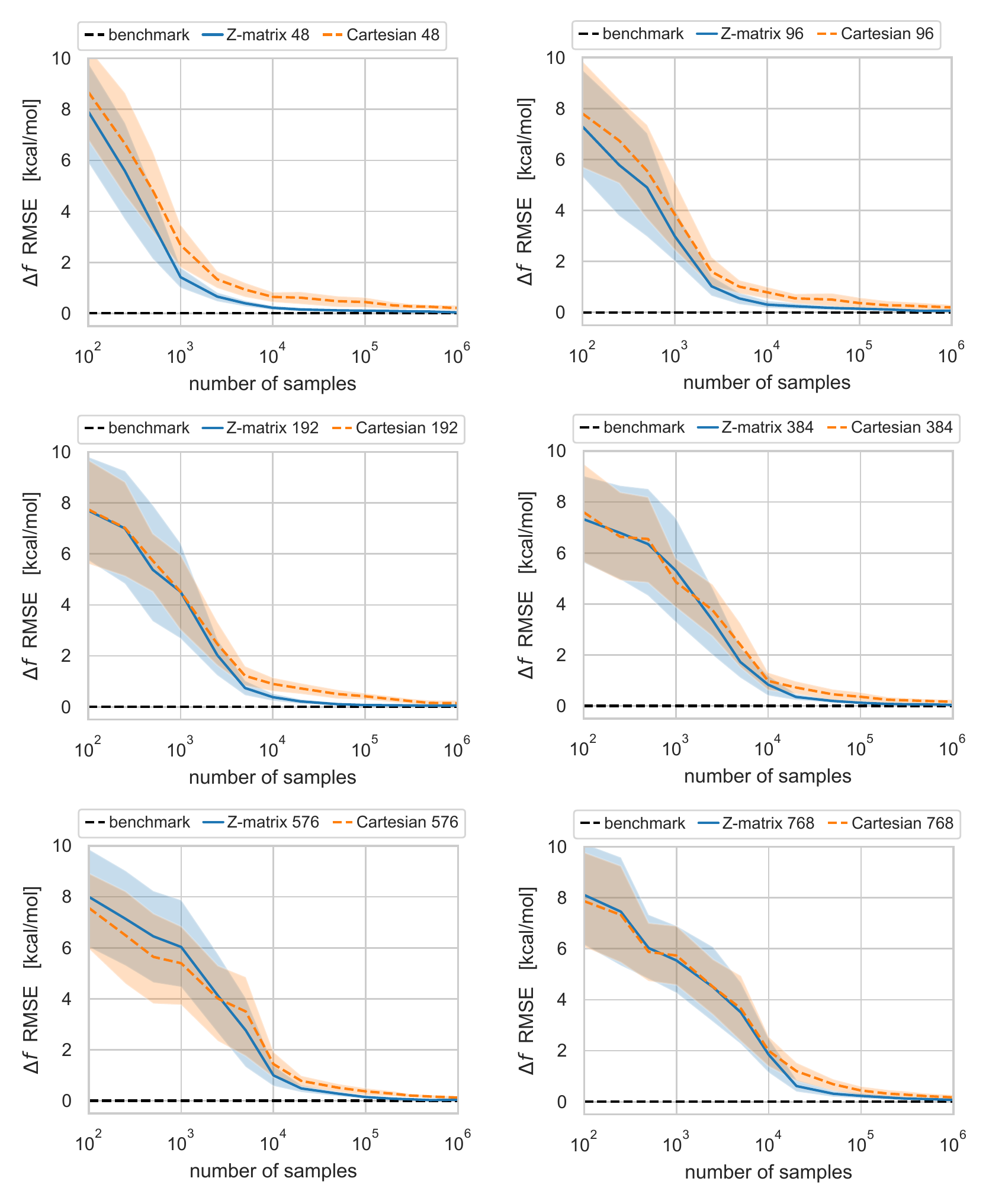}
    \centering
    \caption{\label{fig:si:cartesian-vs-zmatrix}Comparison of the root mean squared error (computed against the benchmark calculations) of the multimap TFEP predictions for the ``good" set using Z-matrix (solid orange) or Cartesian (dashed blue) coordinates for different batch sizes: 48, 96, 192, 384, 576, and 768.}
\end{figure}

\begin{figure}[!h]
    \includegraphics[width=0.9\textwidth]{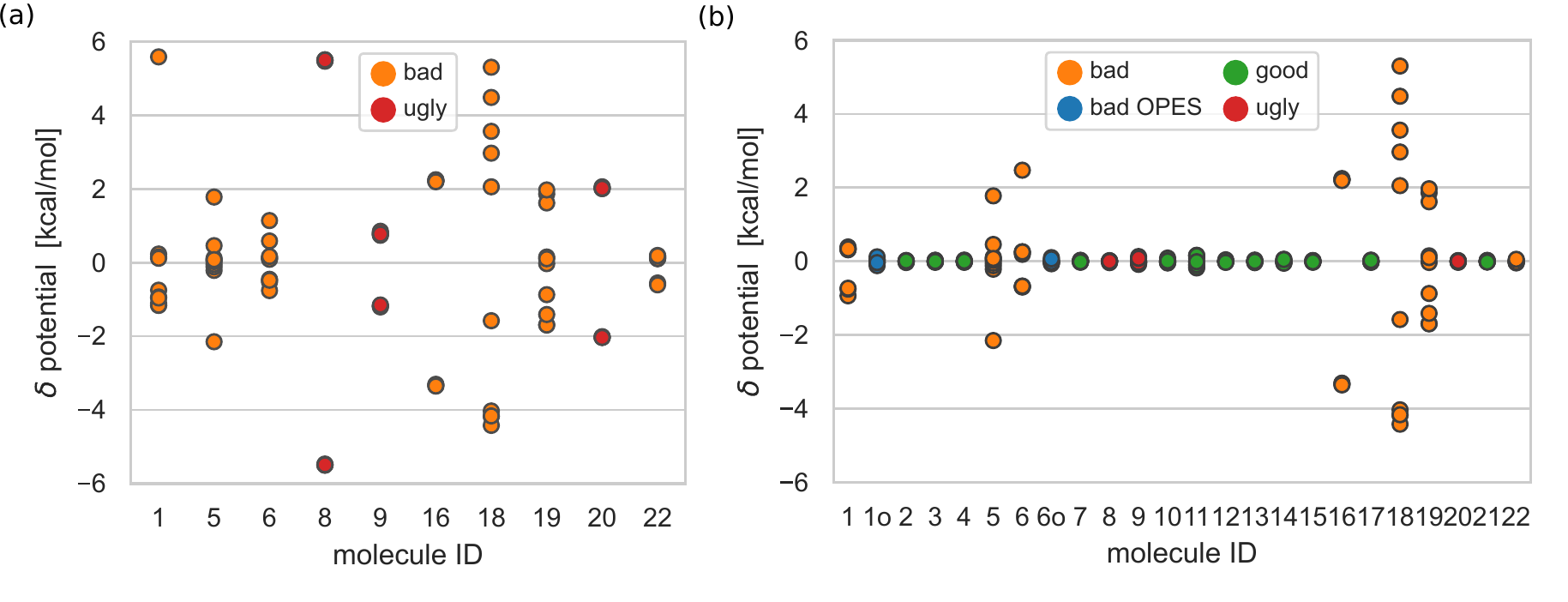}
    \centering
    \caption{\label{fig:si:avg-potential-repeats}Deviation of the potential energy of the 10 replicate simulations (or from 10 different of the OPES simulation) using the dihedral randomization protocol that isomerized the double bonds (a) or not (b).}
\end{figure}

\begin{figure}[!h]
    \centering
    \includegraphics[width=0.9\textwidth]{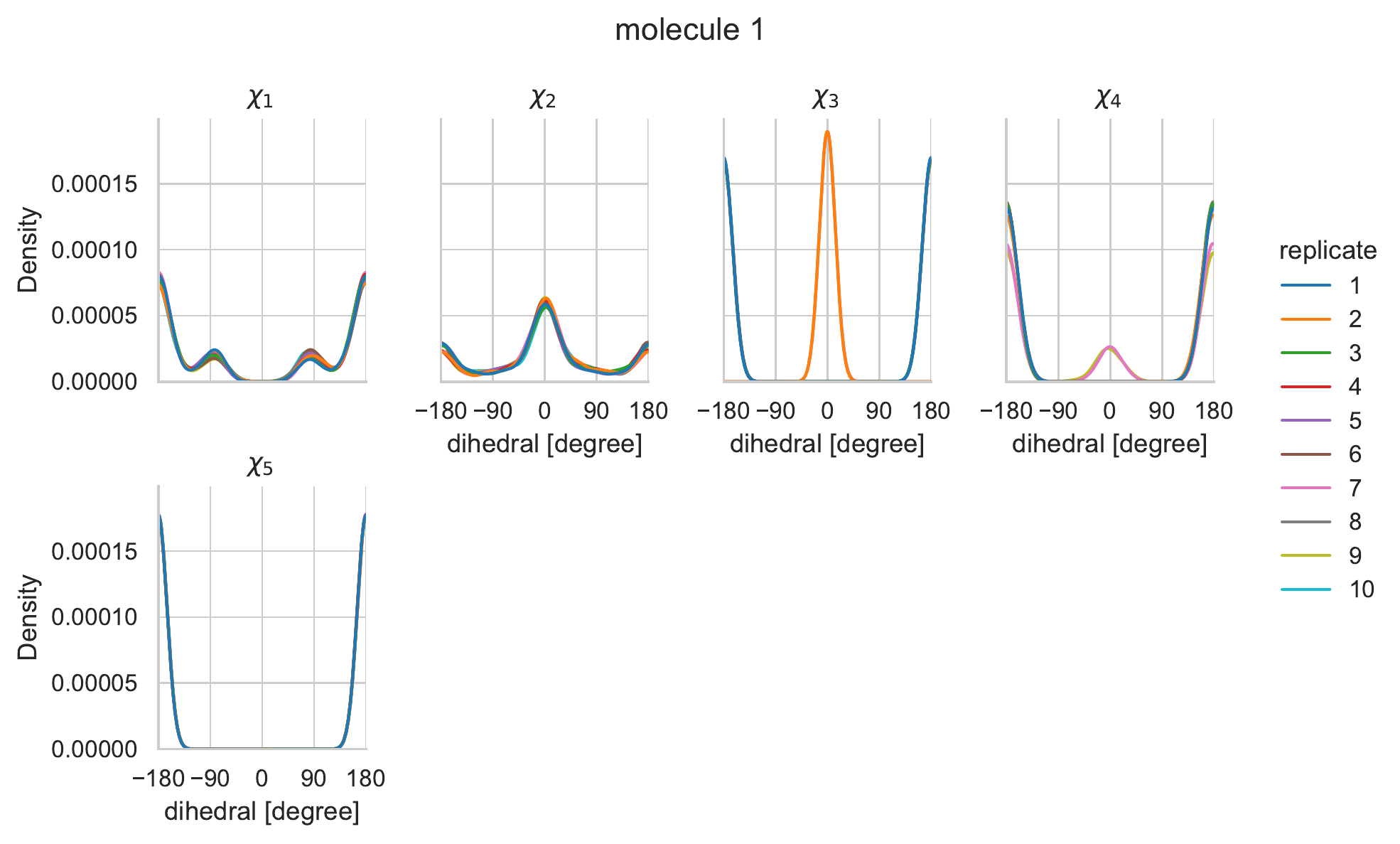}
    \includegraphics[width=0.9\textwidth]{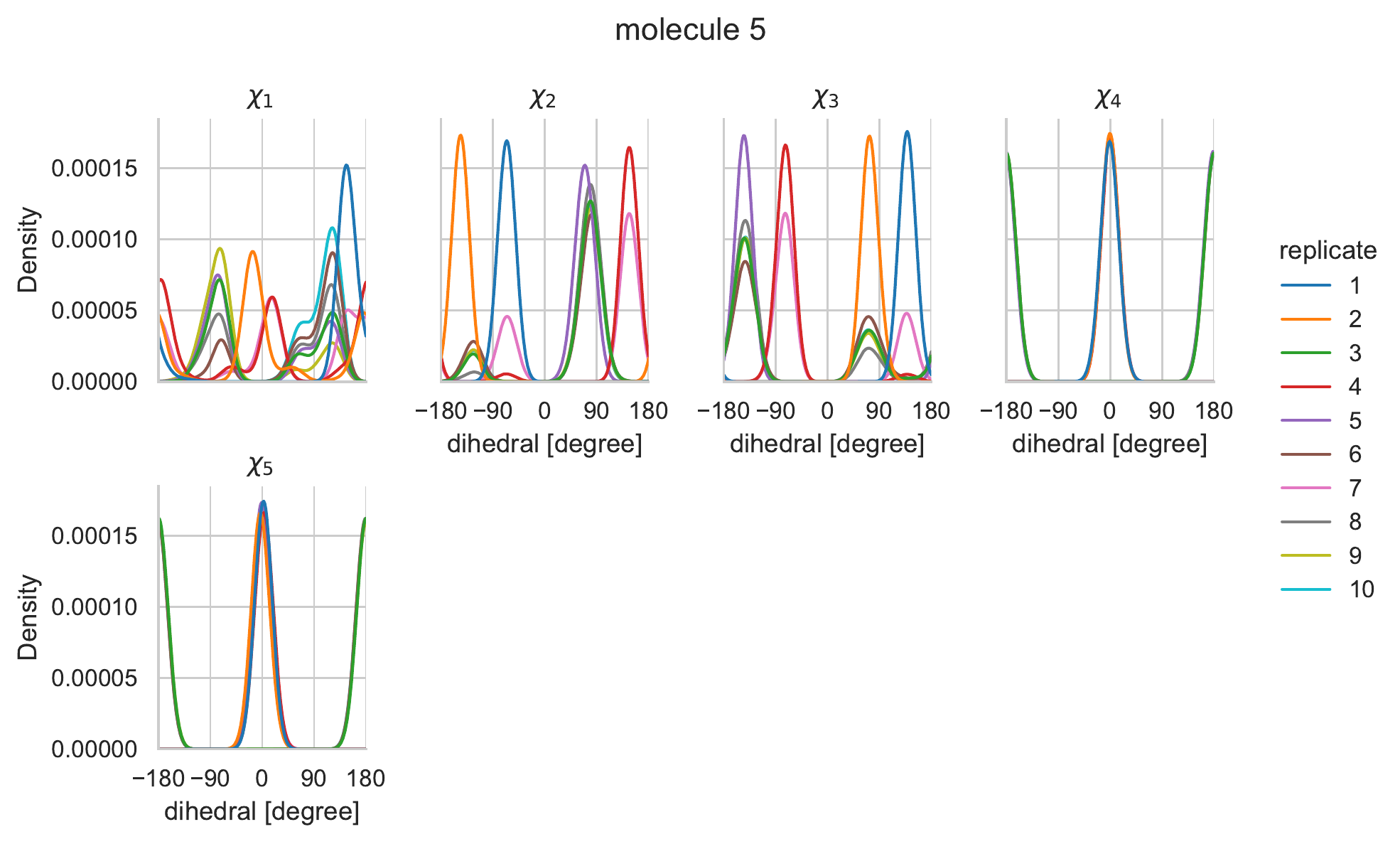}
    \includegraphics[width=0.9\textwidth]{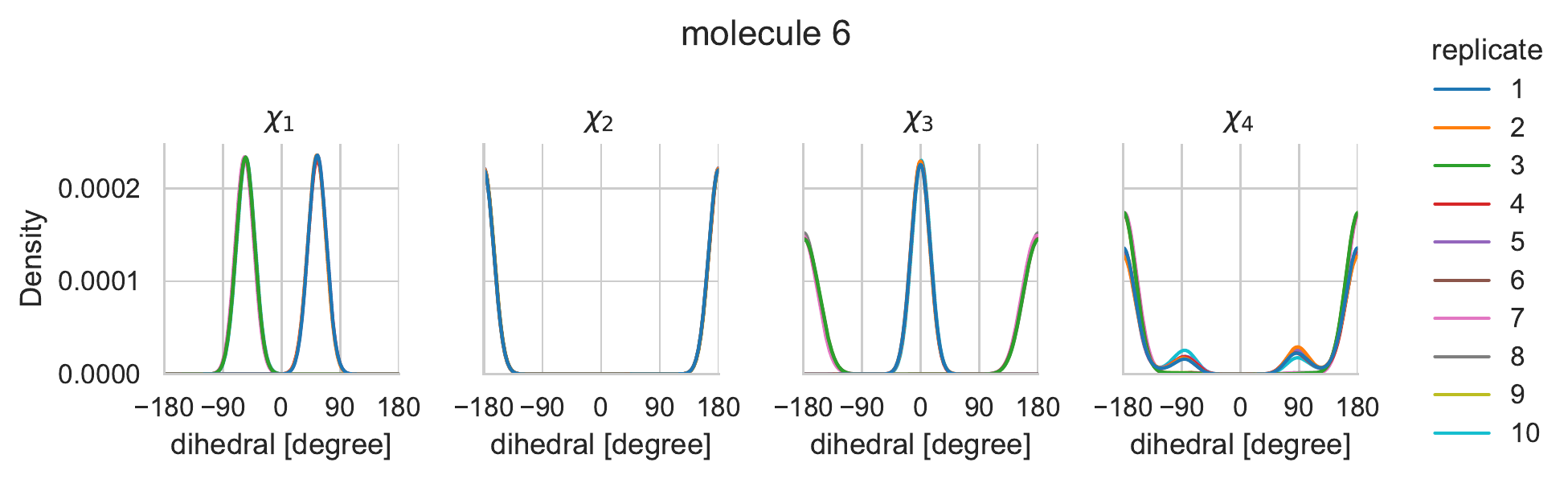}
    \caption{\label{fig:si:dihedrals-bad}Distributions of the dihedrals sampled by the 10 independent replicate simulations for the ``bad" and ``ugly" set.}
\end{figure}
\begin{figure}[!h]
    \ContinuedFloat
    \centering
    \includegraphics[width=0.9\textwidth]{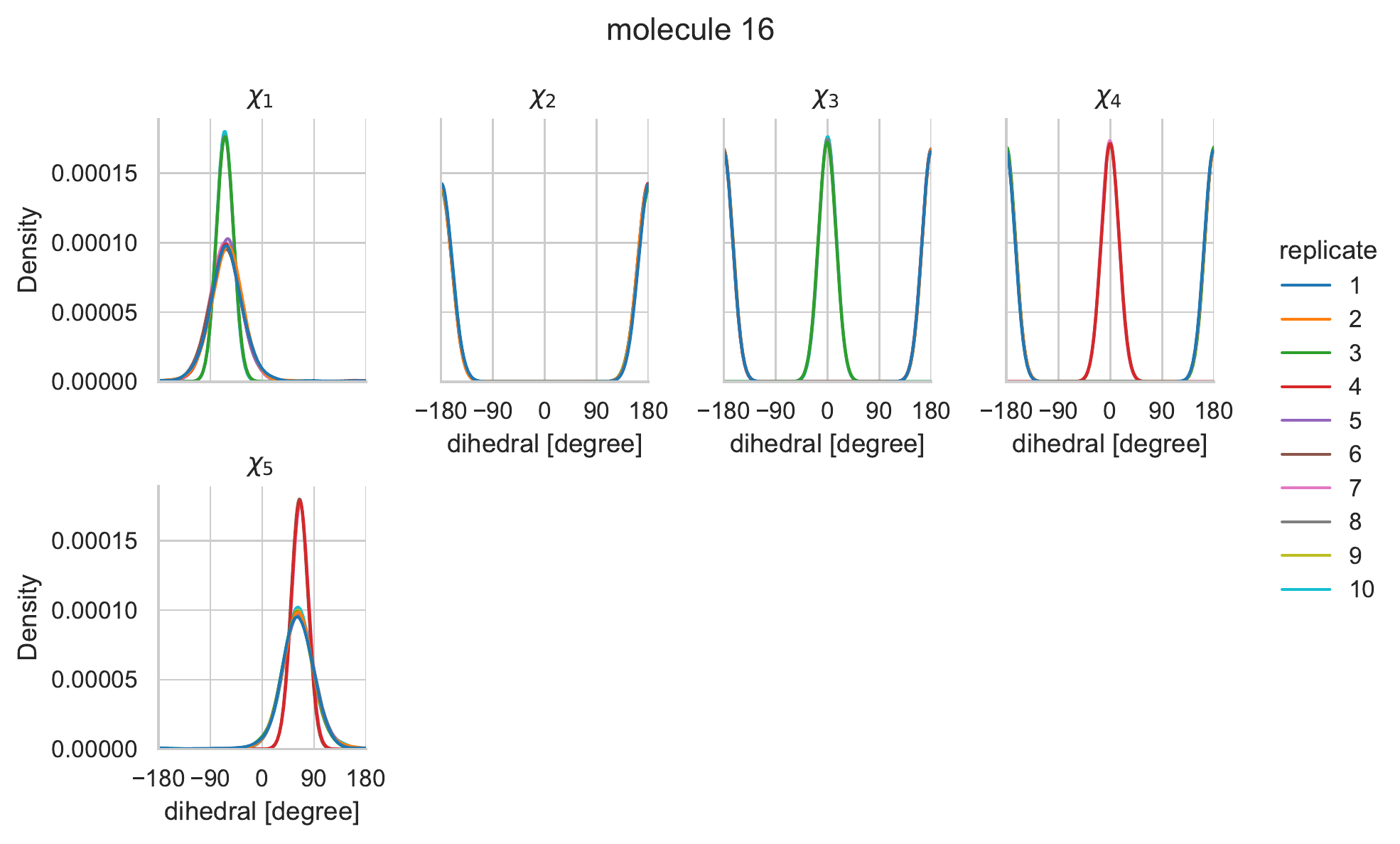}
    \includegraphics[width=0.9\textwidth]{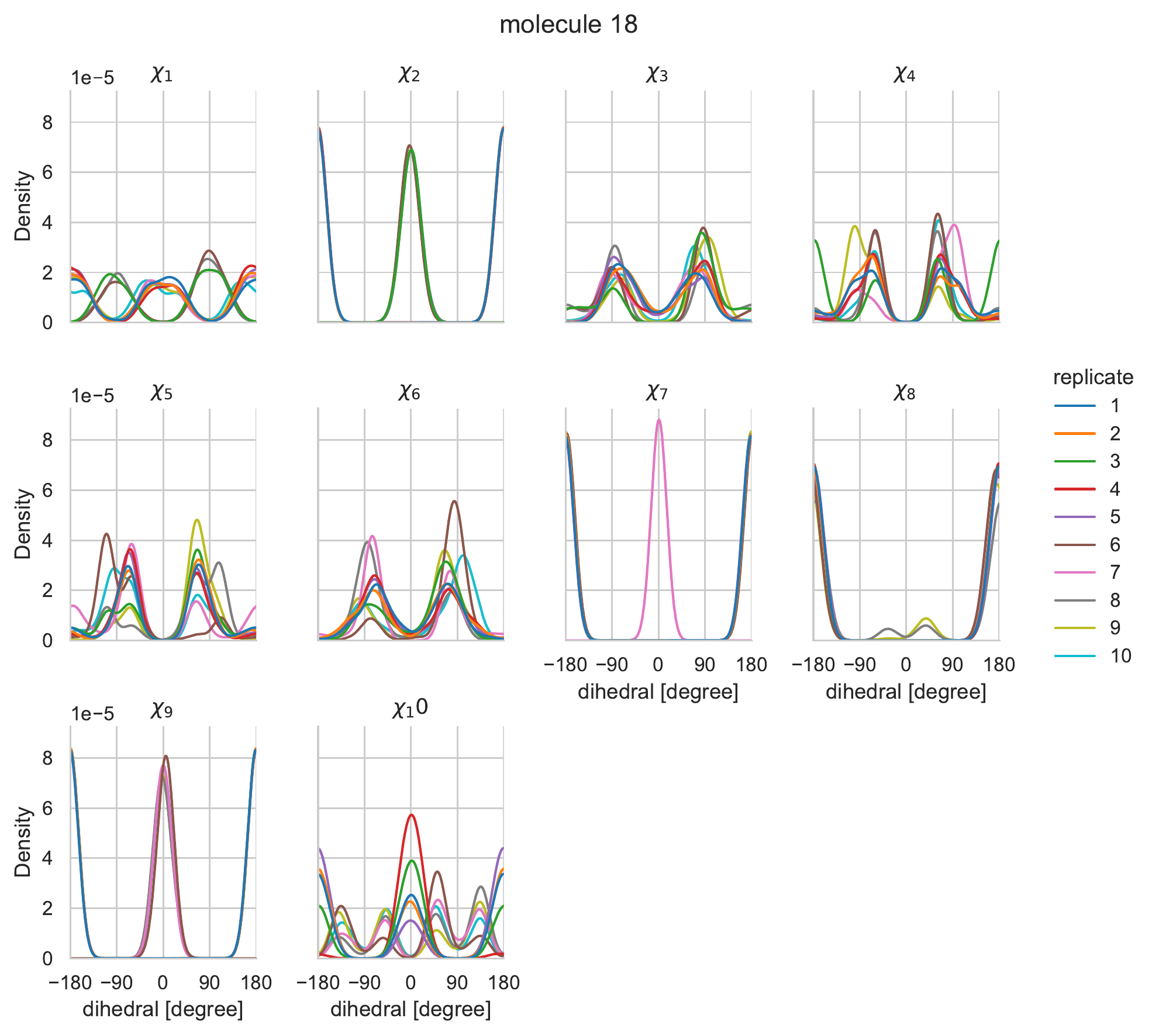}
    \caption{cont.}
\end{figure}
\begin{figure}[!h]
    \ContinuedFloat
    \centering
    \includegraphics[width=0.9\textwidth]{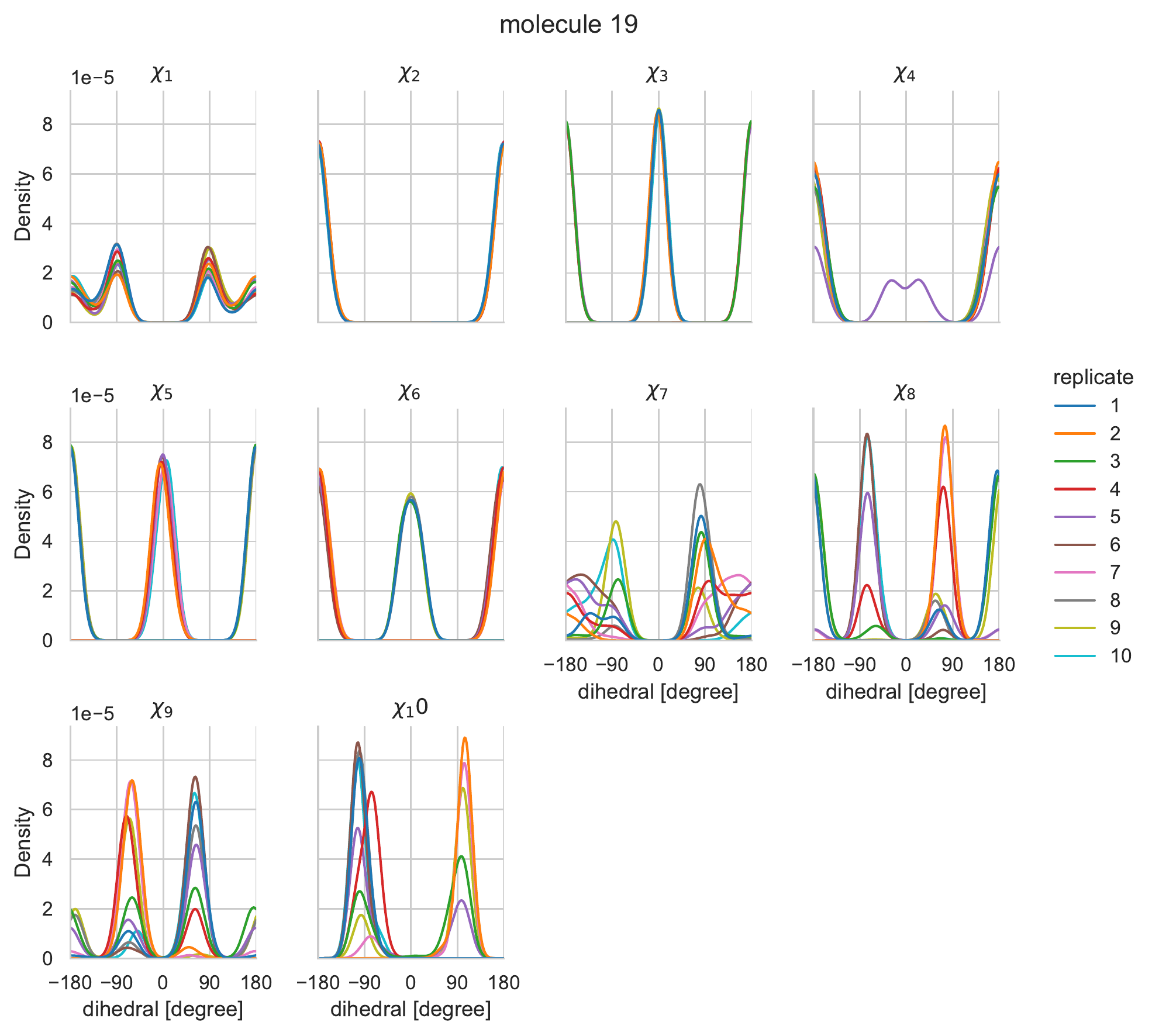}
    \includegraphics[width=0.9\textwidth]{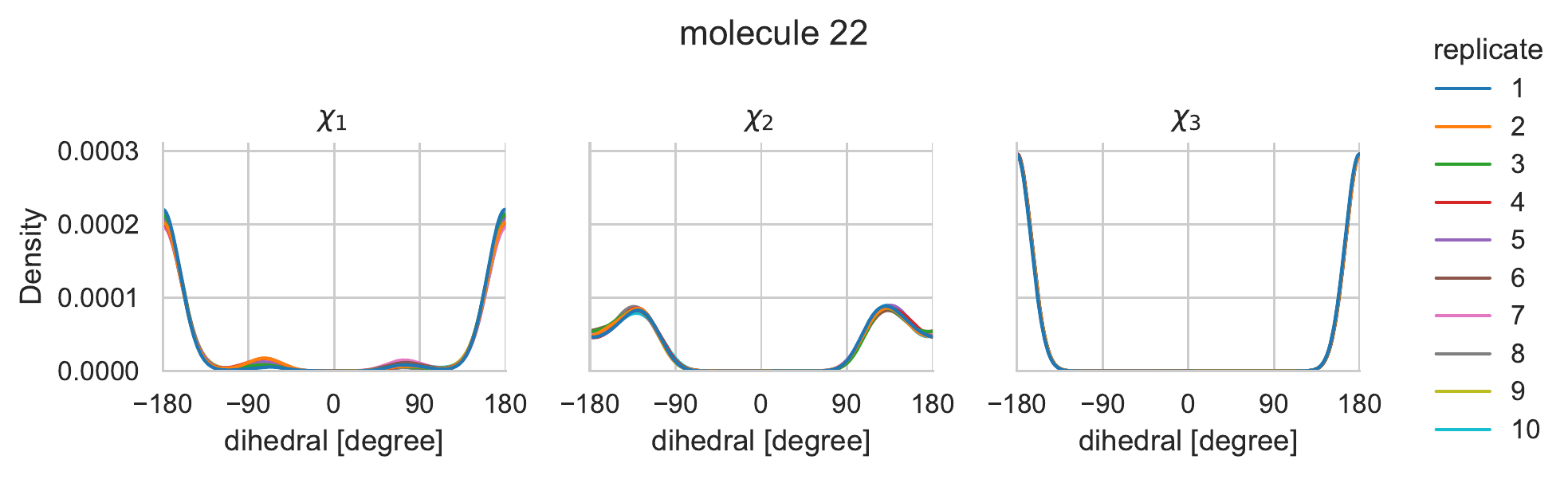}
    \caption{cont.}
\end{figure}
\begin{figure}[!h]
    \ContinuedFloat
    \centering
    \includegraphics[width=0.9\textwidth]{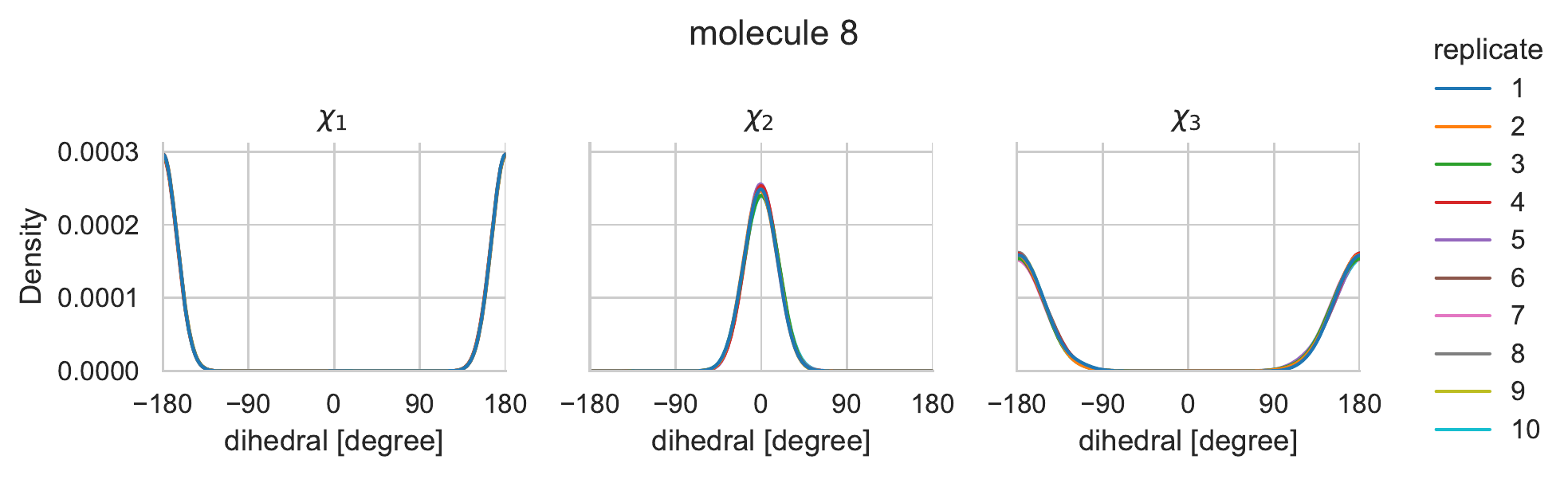}
    \includegraphics[width=0.9\textwidth]{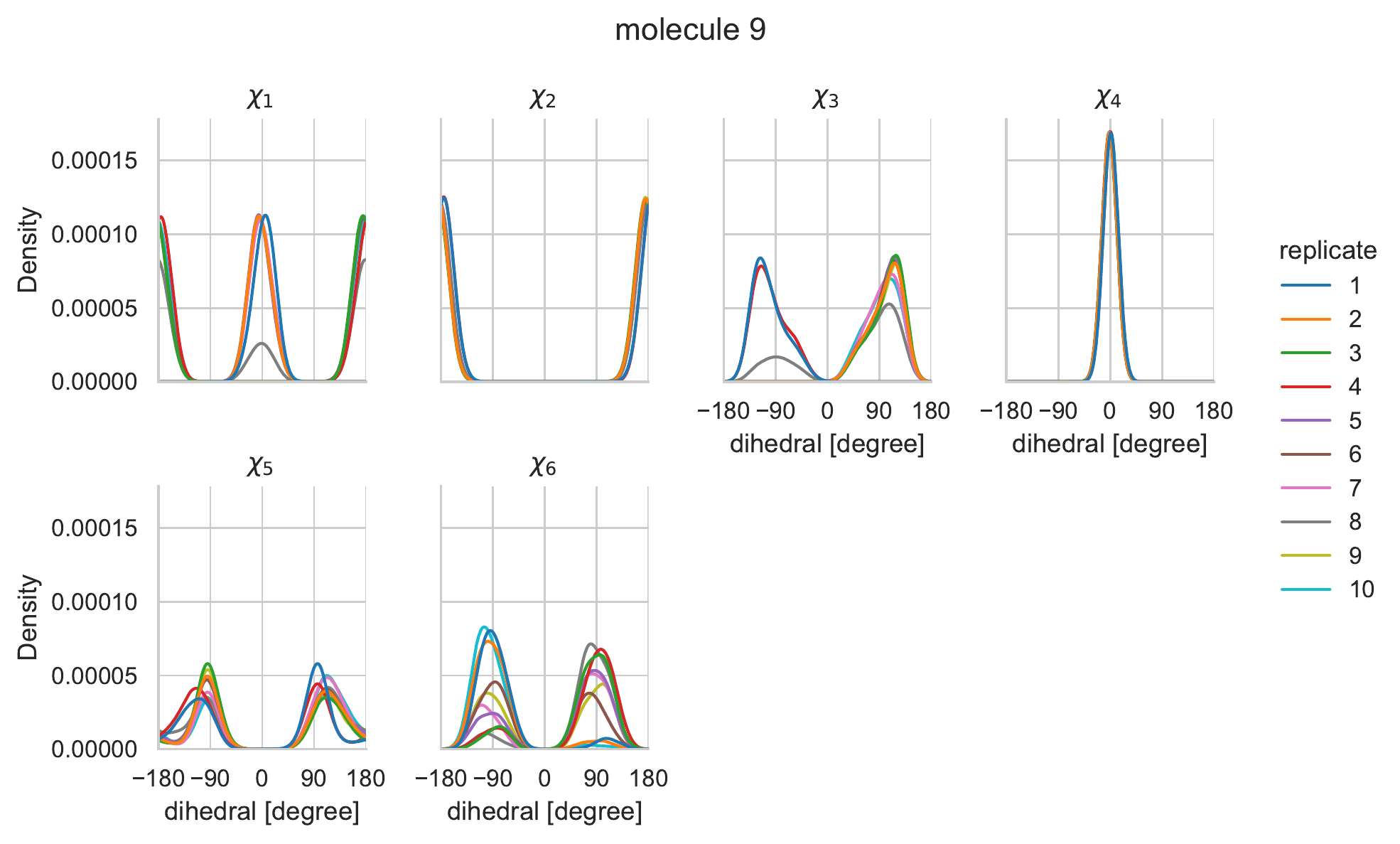}
    \includegraphics[width=0.9\textwidth]{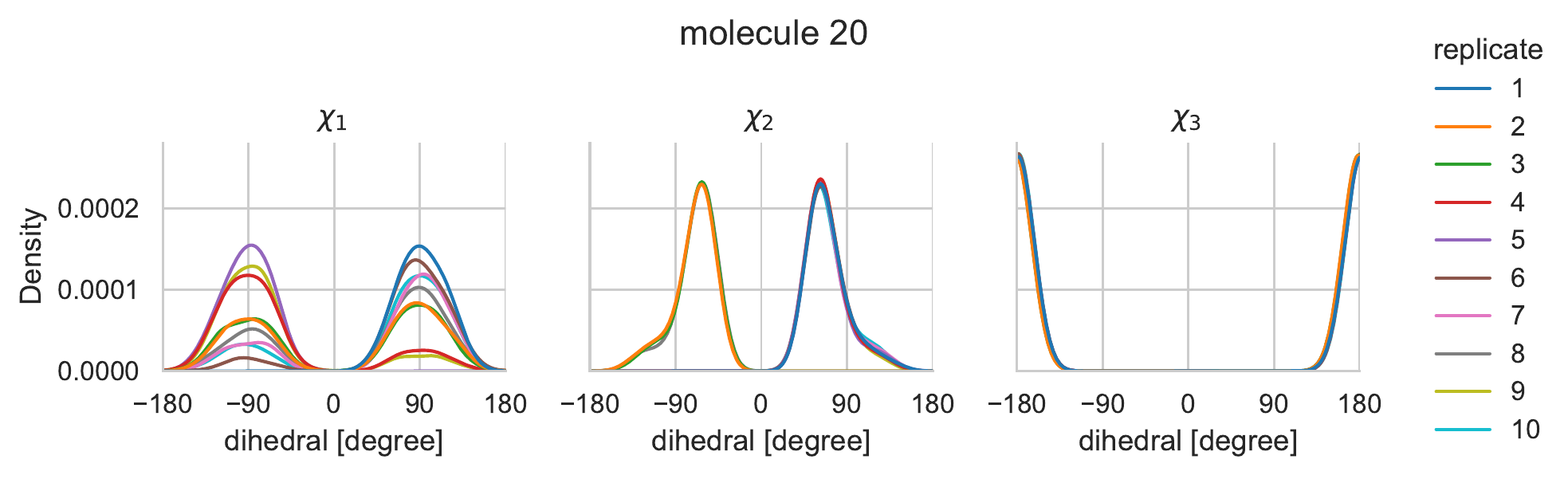}
    \caption{cont.}
\end{figure}

\begin{figure}[!h]
    \centering
    \includegraphics[width=0.9\textwidth]{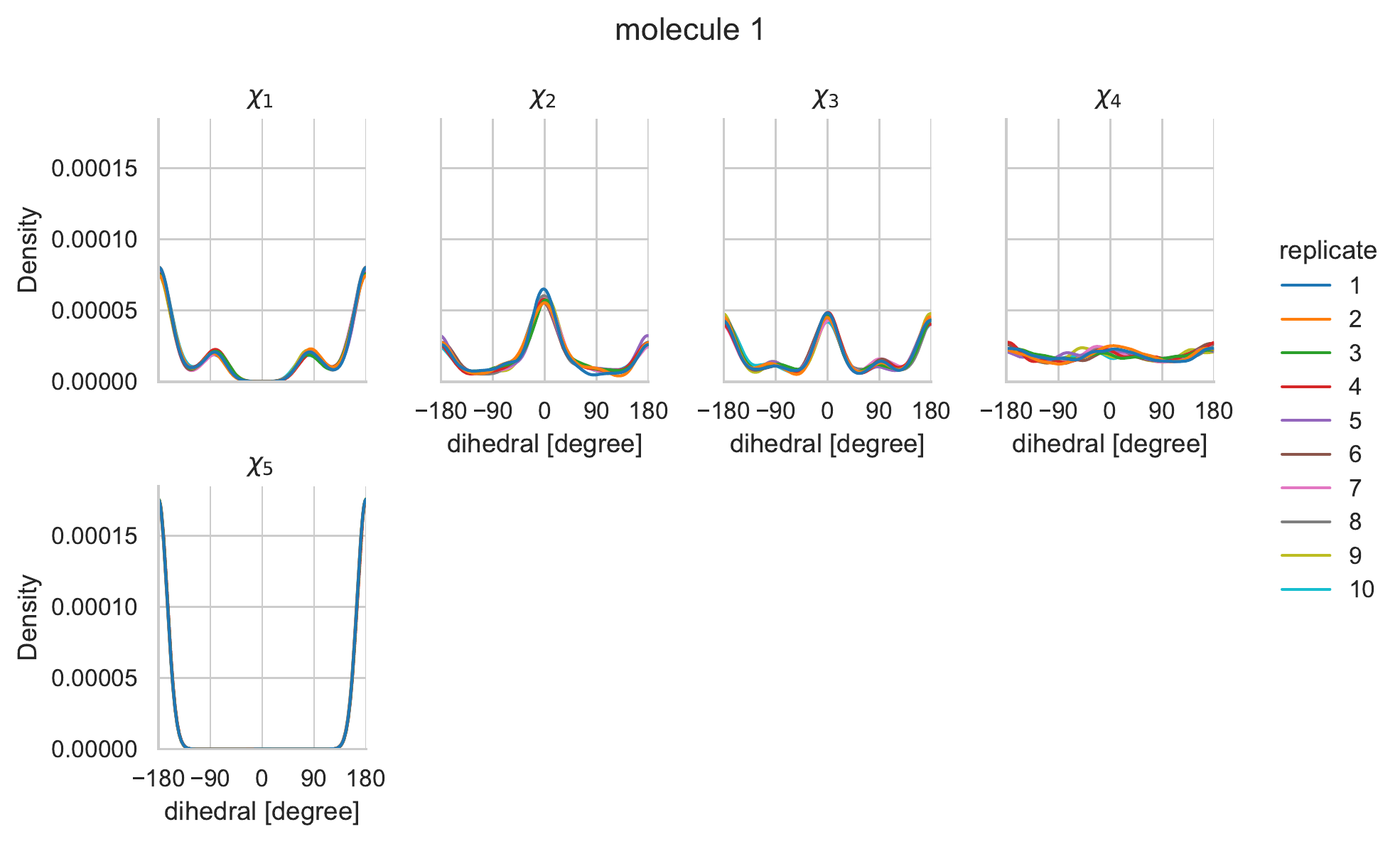}
    \includegraphics[width=0.9\textwidth]{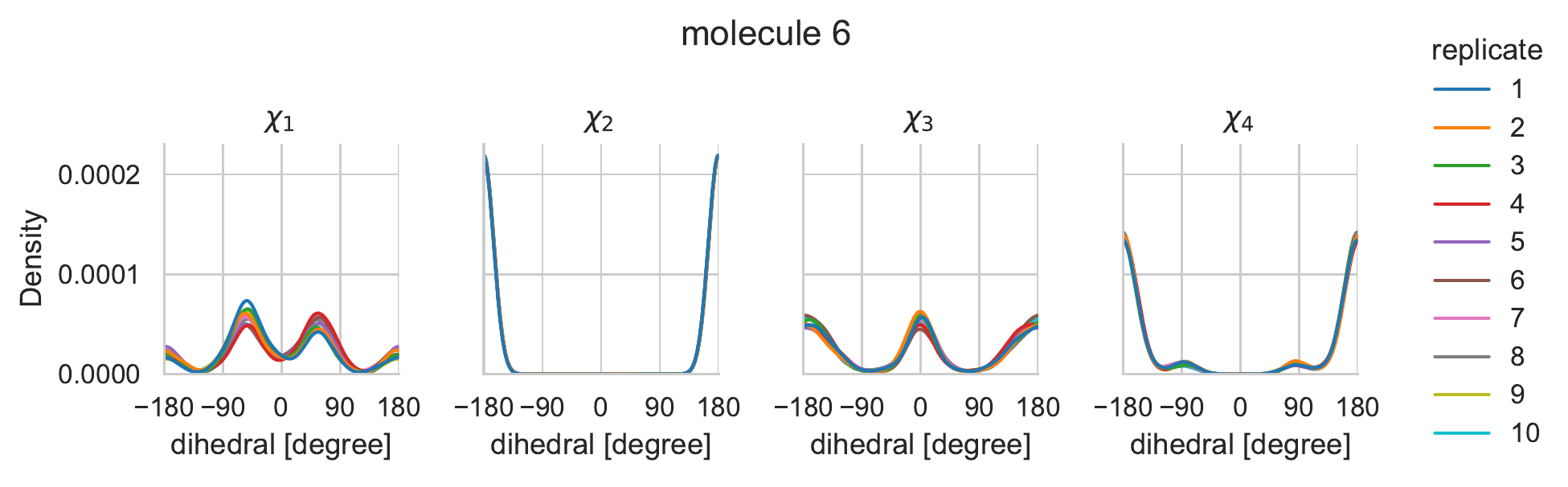}
    \caption{\label{fig:si:dihedrals-opes}Distributions of the dihedrals sampled by the 10 independent replicate simulations for the OPES simulations of molecules \textbf{1} and \textbf{6}.}
\end{figure}

\begin{figure}[!h]
    \centering
    \includegraphics[width=0.9\textwidth]{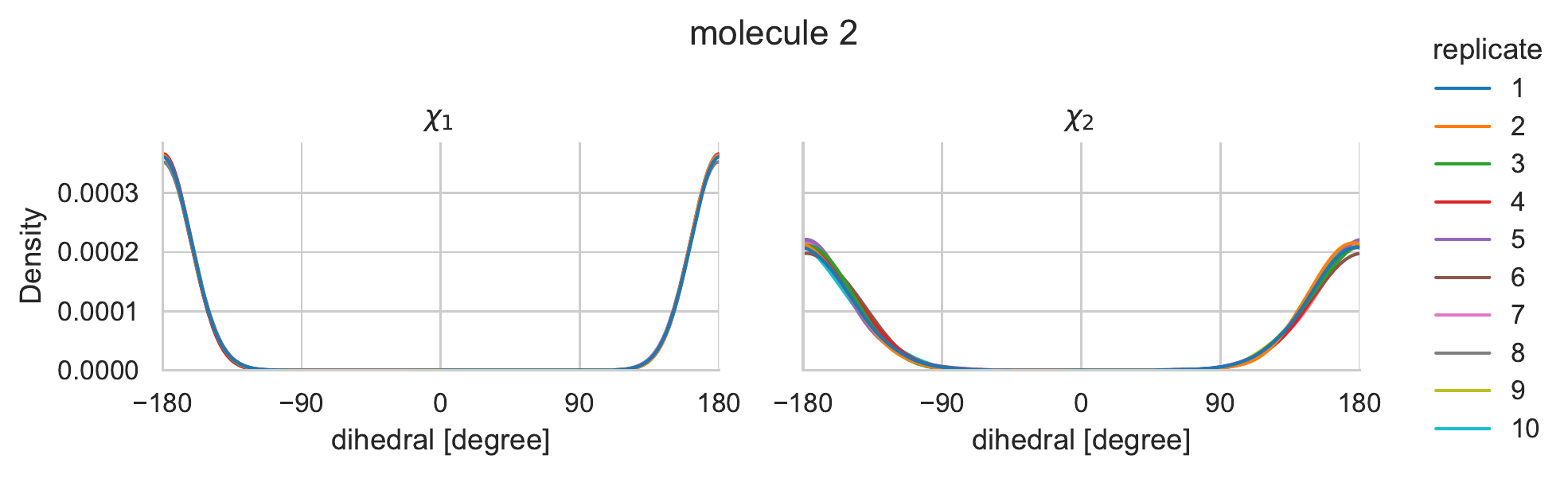}
    \includegraphics[width=0.9\textwidth]{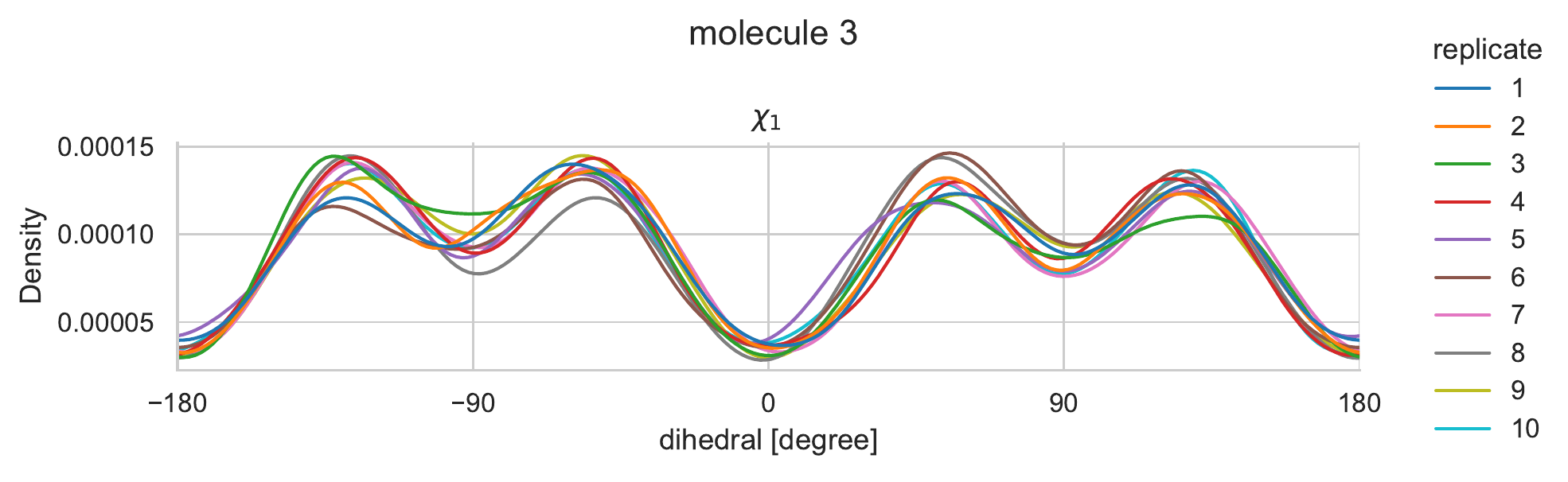}
    \includegraphics[width=0.9\textwidth]{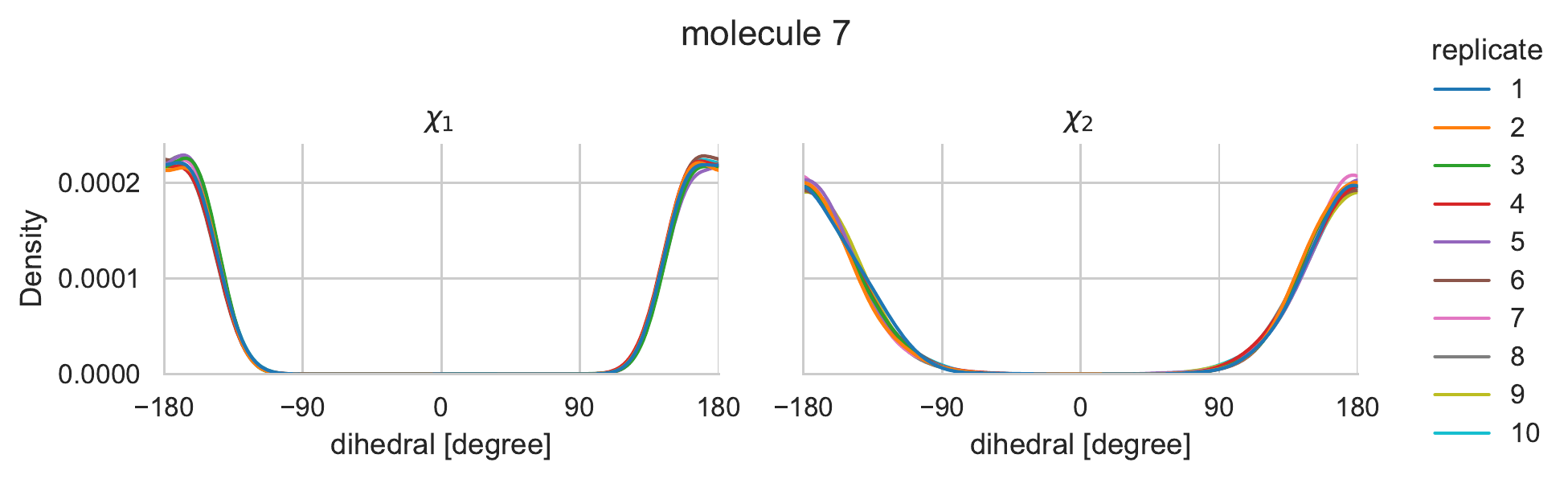}
    \includegraphics[width=0.9\textwidth]{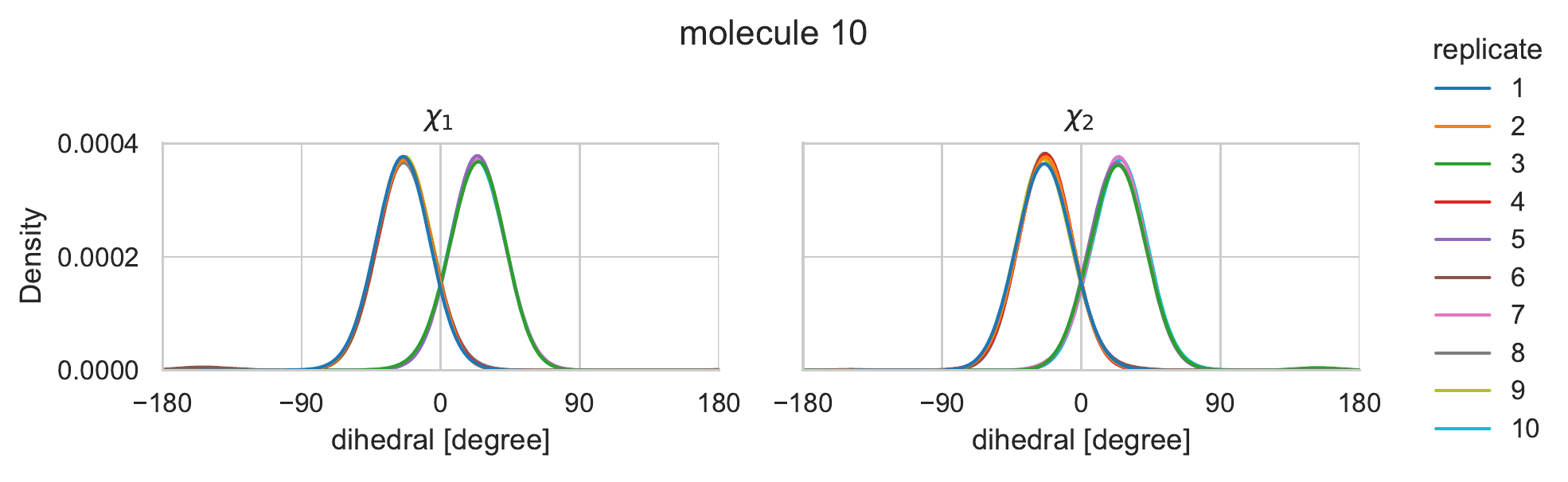}
    \caption{\label{fig:si:dihedrals-good}Distributions of the dihedrals sampled by the 10 independent replicate simulations for the ``good" set.}
\end{figure}
\begin{figure}[!h]
    \ContinuedFloat
    \centering
    \includegraphics[width=0.9\textwidth]{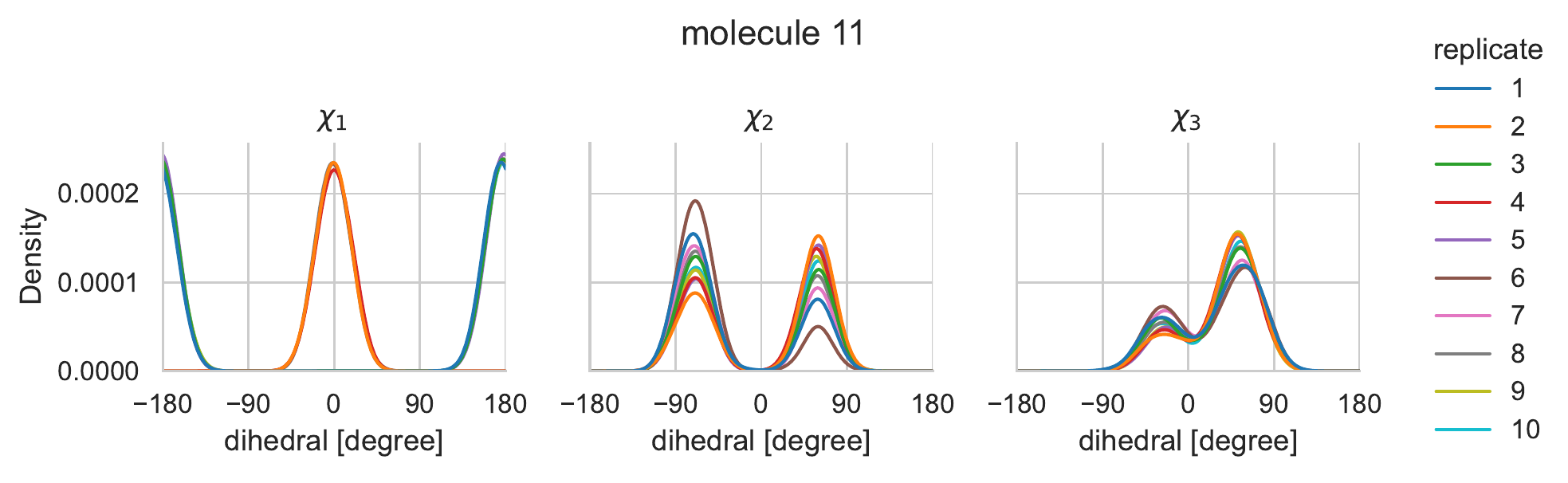}
    \includegraphics[width=0.9\textwidth]{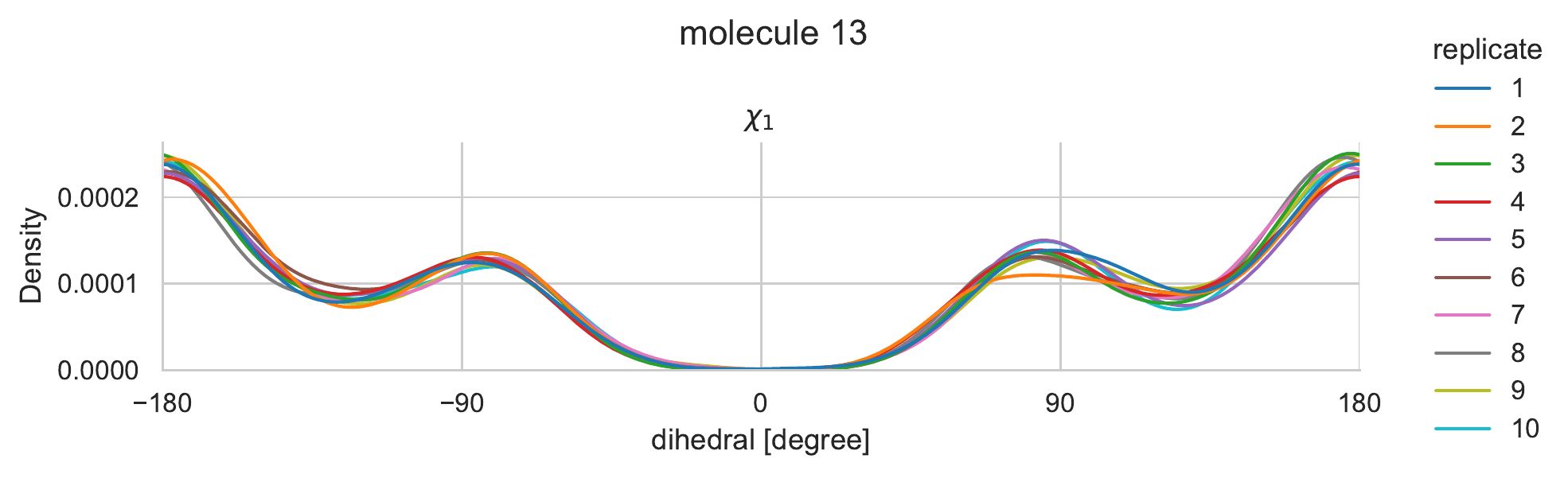}
    \includegraphics[width=0.9\textwidth]{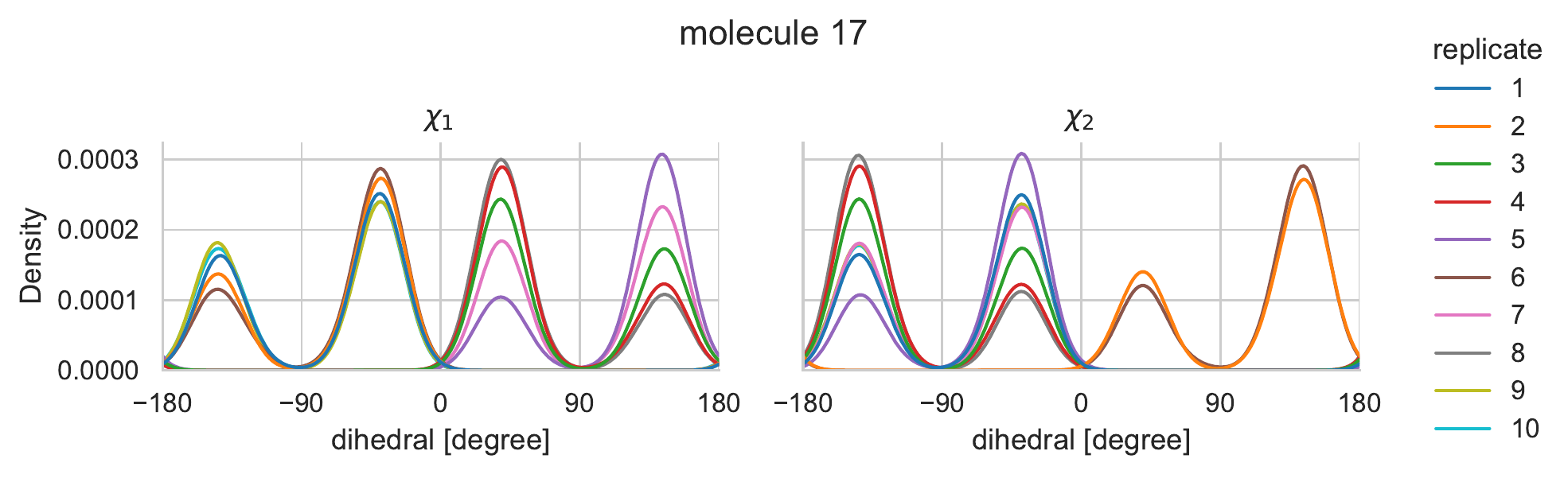}
    \includegraphics[width=0.9\textwidth]{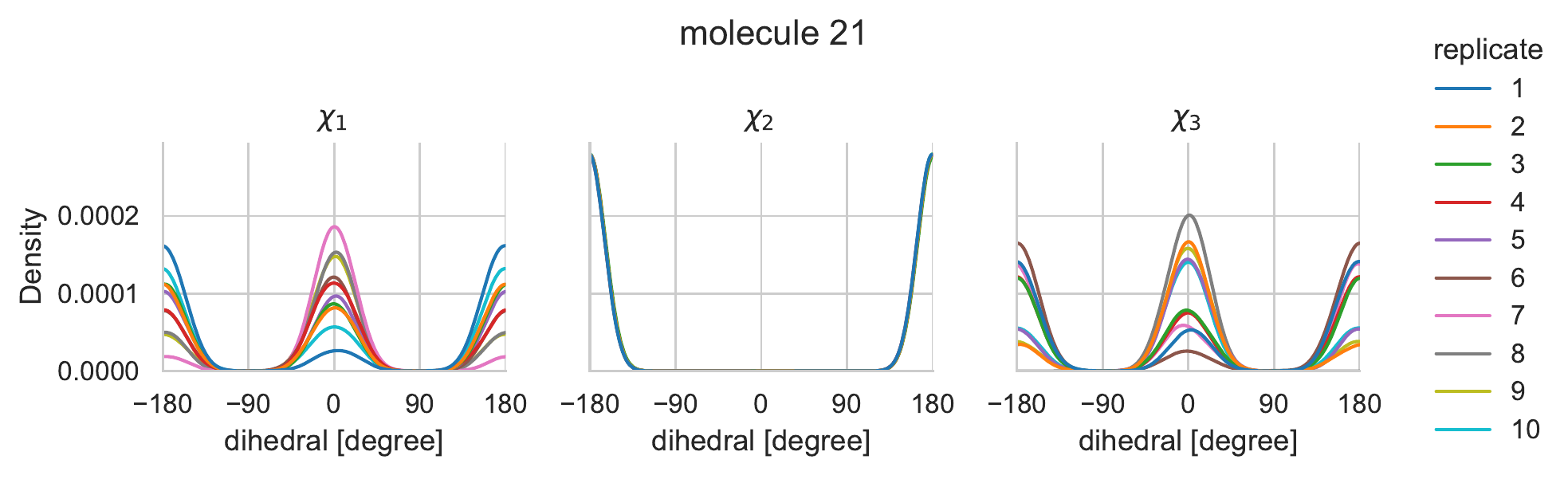}
    \caption{cont.}
\end{figure}

\end{document}